\UseRawInputEncoding
\documentclass[prl,twocolumn,twocolumngrid]{revtex4}

\newcommand{\withpdf}[1]{}

\usepackage{graphicx}
\usepackage{amsfonts}
\usepackage{amsmath}
\usepackage{bm}
\usepackage{alltt}
\usepackage{fancyhdr}

\begin{document}

\title{Radical Anion Functionalization of Two-Dimensional Materials             
as a Means of Engineering Simultaneously High Electronic and Ionic Conductivity Solids
}

\author{K\'aroly N\'emeth$^{\ast}$}
\affiliation{Physics Department, Illinois Institute of Technology, Chicago, Illinois 60616,
USA \\ \email{nemeth@agni.phys.iit.edu} }

\begin{abstract}
A radical anion based functionalization of the basal plane of hexagonal boron nitride (h-BN)
and other two-dimensional (2D) materials is proposed in the present study. 
The resulting materials can 
reversibly be oxidized without the detachment of the functional groups from the basal plane
and can thus serve as surface-intercalation type cathode electroactive species and fast solid ion
conductors at the same time. 
The functionalization of h-BN with [$\cdot$OBX$_{3}$]$^{-}$
radical anions (X=F,Cl) in the presence of Li, Na or Mg cations provides one example of such
systems. This material can be realized in a proposed simple, two step synthesis. In the first
step, a symmetric Lewis adduct of the
corresponding Li, Na or Mg peroxides is formed with BX$_{3}$. In the second step, the anion of
the Lewis adduct is thermally split into two identical 
[$\cdot$OBX$_{3}$]$^{-}$ radical anions that covalently functionalize the B atoms of h-BN.
In the maximum density surface packing functionalization, the product of the synthesis is
A$_{n}$[(BN)$_{2}$OBX$_{3}$] (A=Li,Na with n=1 or A=Mg with n=0.5).
In the highly oxidized states (0$\le$n$\le$1 for Li and Na and 0$\le$n$\le$0.5 for Mg), 
the electronic conductivity of this material is 
in the order of 1 S/cm, similar to carbon black. In the fully reduced states (n=2 for Li and
Na and n=1 for Mg), the material becomes an insulator, like h-BN.
The tunability of the electronic properties of A$_{n}$[(BN)$_{2}$OBX$_{3}$]
via the cation concentration allows for its application as multifunctional material
in energy storage devices, simultaneously serving as cathode active species, solid electrolyte,
electroconductive additive, separator, heat conductor and coating for metal anodes 
that enables dendrite-free plating. This multifunctionality reduces the number of phases needed
in an all-solid-state battery or supercapacitor and thus reduces the interfacial impedance
making energy storage devices more efficient.
For example, Li[(BN)$_{2}$OBF$_{3}$] is predicted to have 5.6 V open circuit voltage vs Li
metal anode, capacity of 191 mAh/g, specific energy of 1067 Wh/kg and can store energy at a
(materials only) cost of 24 USD/kWh.       
\end{abstract}


\maketitle

\section{Introduction}
The quest for high performance electrochemical energy storage devices poses
a multi-objective optimization problem over the space of readily available and potentially
synthesizable designer materials.                     
The objectives include high energy and power density, safety of operation, economic
and environmentally friendly composition, rechargeability, fast charging and
discharging, a long cycle-life (i.e. large number of charge/discharge cycles at a
steady performance) and reliable functioning in a broad temperature window.

Two-dimensional (2D) layered materials represent an attractive platform for 
developing effective rechargeable 
intercalation cathodes for these objectives as they allow for the
reversible intercalation of Li$^{+}$ ions between the layers. 
This principle was first demonstrated on TiS$_{2}$ 
\cite{whittingham1976electrical} while a more practical approach using Li$_{x}$CoO$_{2}$
(0$<$x$<$1) was developed at the same time \cite{godshall1980thermodynamic,KMizusima80}.
Since then, variants of Li$_{x}$CoO$_{2}$ where Co is partially
substituted by Ni, Mn and Al have also been developed to mitigate the scarcity of cobalt  
\cite{MMThackeray01A,MMThackeray01B,JRDahn01,chen2004aluminum}. 

On the anode side, graphite, another 2D layered material, and similar carbonaceous
materials containing layered fused aromatic rings (such as coke) 
were proposed and successfully 
applied as Li-intercalation systems \cite{AYoshino86,RYazami83}. 

Electrode materials can be divided into two groups: intercalation based ones (such
as the LiCoO$_{2}$ cathode and the graphite anode) and
conversion based ones (such as the sulfur cathode \cite{chen2014recent,pang2019lightweight} 
and the silicon anode \cite{ashuri2016silicon}). 
Whenever possible, intercalation based electrodes appear
more attractive as they typically have much smaller
volume changes than conversion ones during the cycling of the battery.
Intercalation based electrodes also have a more steady chemical structure as
compared to conversion based ones. Chemical bonds in the latter ones may go
through radical transformation during cycling.

The present paper is highly interdisciplinary as it combines 
fields of theoretical materials design, quantum chemical prediction of properties, 
synthesis design, thermal stability analysis, electrochemistry and economic analysis.
Therefore, it is important to discuss its layout and logic first, before the
elaboration starts.      

First, a thorough review is
provided about past and present attempts to achieve the objectives mentioned at the
beginning of the present Introduction. Particular attention is given to 
designer strategies on 
how simultaneous high energy and high power density batteries might be realized.
This review describes the historical roots and emergence of the graphene oxide (GO) 
battery as the sole existing demonstration of high energy and high power energy storage at 
the same time. 

We will point out that the great performance of the GO battery is conditional to the
presence of a Lewis acid donating electrolyte and this Lewis acid forms Lewis adduct
with GO. A particularly important Lewis acid is boron trifluoride (BF$_{3}$) donated
by (Li/Na)BF$_{4}$ electrolyte. The Lewis adduct involves various forms of 
the -OBF$_{3}$ functional group that plays a key role in the present study.
The review also discusses fundamental problems of the GO problem,
especially its safety problem attributed to explosive decomposition of GO upon
heating. 

At this point, a reference will be made to our former works 
\cite{nemeth2015functionalized,nemeth2020-US10693137,nemeth2014materials,nemeth2018simultaneous} 
in which we proposed to substitute GO with functionalized hexagonal boron nitride (h-BN) 
in order to cure the thermal stability problem of GO.
In addition, we have predicted that the -OBF$_{3}$ functionalization of h-BN would be
particularly advantageous for an efficient, h-BN-based battery. This makes the key
importance of the -OBF$_{3}$ functional group very clear, in both GO and h-BN-based
batteries.

A question naturally emerges about whether such an -OBF$_{3}$
functionalization of h-BN is possible at all. We answer this question first by
reviewing the literature for analogous functionalized h-BN. It turns out that the
historically first functionalized h-BN was actually very similar: it contained
-OSO$_{2}$F functional groups on the surface of h-BN, covalently bound to the B atoms.
This material has been synthesized first in 1978 \cite{bartlett1978novel} and its
electrical conductivity was measured and found to be 
surprisingly high, 1.5 S/cm \cite{shen1999intercalation}, especially compared to the
near 6 eV band gap and insulator nature of h-BN. The -OSO$_{2}$F functional groups can
cover the surface of a h-BN monolayer with very high density leading to every second B
atom functionalized in (BN)$_{2}$OSO$_{2}$F.

Unfortunately, (BN)$_{2}$OSO$_{2}$F is not stable when in contact with base metals.
It decomposes irreversibly due to the evolution of volatile
molecules. Therefore, (BN)$_{2}$OSO$_{2}$F cannot be used directly in batteries.
However, the synthesis of (BN)$_{2}$OSO$_{2}$F provides a great model for
the synthesis of our proposed -OBF$_{3}$ functionalized h-BN, which has the
formula of (BN)$_{2}$OBF$_{3}$ in the most dense surface packing.

At this point, we calculate and analyze the
electronic structure of the compounds mentioned so far, i.e. h-BN,
(BN)$_{2}$OSO$_{2}$F and (BN)$_{2}$OBF$_{3}$ as well as the reduced versions of
(BN)$_{2}$OBF$_{3}$ that form when the (BN)$_{2}$OBF$_{3}$ is in contact 
with Li, Na or Mg. Furthermore, the same
analysis is also carried out for (BN)$_{2}$OBCl$_{3}$ which contains Cl instead of F
in the same structure. This analysis establishes the stability of the reduced versions
of (BN)$_{2}$OBF$_{3}$ as opposed to (BN)$_{2}$OSO$_{2}$F. It also establishes the
similarity between the band structures, electronic conductivities and magnetic
properties of these materials.

Following the quantum chemical analysis, we analyze the stability and decomposition
reactions of (BN)$_{2}$OBF$_{3}$ and its reduced versions on the basis of analogy with
the known decomposition of (BN)$_{2}$OSO$_{2}$F. It
turns out that (BN)$_{2}$OBF$_{3}$ has solid decomposition products and therefore it
is much safer than GO or (BN)$_{2}$OSO$_{2}$F.

Next, we utilize the synthesis model provided by
(BN)$_{2}$OSO$_{2}$F and propose a similar, simple synthesis for the reduced versions
of (BN)$_{2}$OBF$_{3}$. Additionally, we also propose and analyze alternative
synthesis methods.  

The above are followed by a discussion of application areas of (BN)$_{2}$OBF$_{3}$.
These include cathode active species, solid electrolytes, electroconductive additives,
separators and anode coatings for dendrite-free plating of Li, Na and Mg.

Finally, an economic analysis is provided for the economic viability of the proposed
batteries.

\section{Design strategies for high energy and high 
power batteries}
\subsection{Artificial intercalation systems}
Artificial intercalation systems may be built by inserting ``pillaring'' ions or
molecules between the loosely bound layers of some 2D materials 
thereby expanding the interlayer distances 
\cite{Pellion-MgIntercalation-2013,Yao-MgIntercalation-2016}.
The interlayer-expanded 2D materials 
allow for the rapid intercalation of Li$^{+}$, but also that of
larger cations, such as Na$^{+}$, K$^{+}$, Mg$^{2+}$ or complex
cations, such as [MgCl]$^+$ \cite{yoo2017fast}. This technique greatly 
increases both the capacity and rate performance of electrodes based on 2D layered
materials \cite{Yao-MgIntercalation-2016,yoo2017fast}. 
Ideally, a small amount of pillaring agent is sufficient for the interlayer
expansion and most of the interlayer space may be utilized for a dense 
packing of intercalated ions.
A significant advantage of such artificial intercalation systems over naturally
occurring ones is the tunability
of the rate performance of the electrodes through the pillaring agents.
Such artificial intercalation systems are increasingly important for 
beyond-Li-ion batteries, as both the number
of naturally occurring minerals for Na or Mg intercalation is more limited than
that of Li intercalating ones and their rate performance at room temperature
(dependent on ionic conductivity) 
is typically low as compared to the targeted performance. Ideal
intercalation systems should also provide good rate performance in a broad
temperature range, starting as low as -60 $^{o}$C and reaching as far as at
least 200 $^{o}$C: the lower value corresponding to harshest winter
conditions and the highest values to safe operations when the environment of
the battery is highly overheated due to fire, or other sources of radiating heat 
\cite{Zimmerman2015,Zimmerman2017,ergen2018,rodrigues2016hexagonal,cheng2019stabilizing,shen2019chemically,rahman2019high}. 
Artificial intercalation systems appear to provide an efficient solution 
for all the above mentioned problems.

Perhaps the oldest form of interlayer expansion of 2D materials
has been discovered more than a hundred years ago, 
when sulfuric acid was utilized as a pillaring
agent for the interlayer expansion of graphite
\cite{staudenmaier1898verfahren}, as a method that accelerates the
oxidation of graphite and the formation of graphene oxide 
\cite{brodie1859xiii,hummers1958preparation,szabo2006evolution,dimiev2016graphene,dreyer2010chemistry}.
About forty years ago, first stage intercalation compounds of h-BN have been synthesized by the
rapid reaction of h-BN with strong oxidizer
bis(fluorosulfonyl)peroxide (FO$_{2}$S-O-O-SO$_{2}$F),
source of (fluorosulfonyl)oxy radicals ($\cdot$OSO$_{2}$F) \cite{bartlett1978novel,shen1999intercalation}. FO$_{2}$S-O-O-SO$_{2}$F
is a liquid at room temperature and is in equilibrium with $\cdot$OSO$_{2}$F
radicals \cite{shen1999intercalation}.
More recently, a simple drying of concentrated H$_{2}$SO$_{4}$, H$_{3}$PO$_{4}$ 
or HClO$_{4}$ over h-BN also resulted in intercalation and 
interlayer expansion of h-BN \cite{kovtyukhova2013reversible}.

For the purpose of Mg$^{2+}$ or MgCl$^{+}$ ion intercalation, interlayer expansion in 
transition metal containing 2D layered materials, such as in TiS$_{2}$ and MoS$_{2}$, has been
carried out electrochemically whereby the pillaring cations included 
imidazolium, pyridinium, ferrocenium,
alkyl-ammonium, pyrrolidinium or piperridinium cations, while the corresponding
anions were Cl$^{-}$, [(CF$_{3}$SO$_{2}$)$_{2}$N]$^{-}$, BF$_{4}^{-}$ , or AlCl$_{x}$R$_{4-x}$$^{-}$
\cite{Yao-MgIntercalation-2016,yoo2017fast,Pellion-MgIntercalation-2013}. Note that
the pillaring salts in the above systems are neither known to form covalent bonds with
the layers of the host material nor participate in redox reactions. 

\subsection{Edge functionalization and the graphene oxide cathode}
Intercalation may not only happen between two monolayers. It can also happen on the
surface of a functionalized monolayer. In this case, the ions
intercalate between the functional groups that are covalently bound to the monolayer.
As the functionalized monolayers are open to the electrolyte on their entire
surface, the rate of intercalation can be even faster than in the case of
interlayer expanded layered materials. Such intercalation is an example of 
supercapacitor-battery hybrides or of pseudocapacitors.
It is however not necessarily advantageous to exfoliate functionalized
monolayers from the first stage intercalation phase, since bending and
folding might occur decreasing the volumetric capacity of such systems. If
the ionic conductivity of the first stage intercalation compounds is
sufficient, there is no need for exfoliation.

To date, the only broadly studied example of rechargeable cathodes with functionalized 2D
redox active species is based on graphene oxide (GO) 
\cite{jang2011graphene,liu2014lithium,kim2014novel,kim2014all}.
In these studies, the reversible redox reactions are based on oxo (=O) groups attached to
edges and defects of graphene. Oxygen atoms bound to the basal plane as epoxy
groups are largely removed by gentle heating,
as their reduction would cause irreversible formation of oxides. Such
GO cathodes, with up to 21-32 w\% oxygen content, 
have been successfully used with Li and Na anodes.
Very high power densities of 4--45 kW/kg (on average) and
specific energies of 100–-500 Wh/kg have been achieved with Li anodes, and
stable capacity of 300--450 mAh/g 
was maintained over 1000 cycles \cite{liu2014lithium}. 
With Na anode, stable capacity at about 150 mAh/g over 300 cycles and energy
densities between 100--500 Wh/kg were achieved, at power densities as high as 
55 kWh/kg \cite{kim2014novel}, when NaBF$_{4}$ electrolyte additive was
used. With NaClO$_{4}$ additive, capacity decreased from 150 to 50 mAh/g in
300 cycles,
indicating potential reactions between GO and the electrolyte additives.

While the incorporation of boron in the surface of GO was pointed out in Ref
\onlinecite{kim2014novel}, its precise mechanism has not been described yet.
F$^{-}$ ions of BF$_{4}^{-}$ can be exchanged to
oxide ions to some extent, in the present case with oxide ions bound to the
surface or edge of graphene.
Similar reactions were observed for example in the reaction of
hydrogen peroxide with KBF$_{4}$ \cite{petrenko1902}, 
in the reaction of BF$_{3}$ with GO \cite{samanta2013highly} or in the  
formation of perborates \cite{koga2018multistep}, 
also see discussion below for cyclic perborates.
Furthermore, surfaces of metal oxides, such as Al$_{2}$O$_{3}$, SiO$_{2}$, TiO$_{2}$,
ZrO$_{2}$, NiO, etc, are also known to irreversibly absorb BF$_{3}$
leading to the formation of (M)-OBF$_{2}$, (M)-OB$_{3}^{-}$ and (M)-O-(BF)-O-(M) 
functional groups on the surfaces of these materials, where M is a metal atom 
\cite{BF3Al2O3Patent1947,rhee1968chemisorption,morrow1976infrared,sadeghi2008bf}. 

Based on the analogy to the reaction of GO and BF$_{3}$ in Ref
\onlinecite{samanta2013highly} and that of oxidized h-BN and BF$_{3}$
as discussed in our former work \cite{nemeth2018simultaneous} the likely mechanism of
boron incorporation from NaBF$_{4}$ 
into GO may be based on the following equations in the discharged state:
\begin{equation}
\rm
G{-}O^{-} + BF_{4}^{-} \rightleftharpoons G{-}O{-}BF_{3}^{-} + F^{-} ,
\end{equation}
where G stands for graphene and G--O$^{-}$ for reduced oxygen atoms covalently bound to graphene.
In the charged state or during charging, the following reactions may happen:
\begin{eqnarray}
\begin{aligned}
\rm
G + G{=}O + BF_{4}^{-} &\rm \rightleftharpoons G{-}O{-}BF_{3}^{-} + G{-}F, \\
\rm
G + G{-}O{-}BF_{3}^{-} &\rm \rightleftharpoons e^{-} + G{-}O{-}BF_{2} + G{-}F, \\
\rm
G{-}O{-}BF_{2} + F^{-} &\rm \rightleftharpoons G{-}O{-}BF_{3}^{-}.
\end{aligned}
\end{eqnarray}
The above reactions are likely responsible for the steady and even somewhat increasing
capacity during cycling
of the GO cathodes with NaBF$_{4}$ electrolytes in Ref \onlinecite{kim2014novel}. 
The increasing capacity may
originate from the utilization of epoxy oxygen of GO that would otherwise irreversibly form
Na$_{2}$O. The conversion of edge-bound oxo groups to their Lewis adducts with BF$_{3}$
may further stabilize these units on the edge of graphene. Thus, the Lewis adducts of
both epoxy and oxo groups of GO may become more stable resulting in stable capacities.
It is very likely, although not noticed yet, 
that the use of LiPF$_{6}$ electrolyte additive leads to 
analogous reactions and may contribute to the cycling stability of Li-GO batteries 
\cite{jang2011graphene,liu2014lithium}, as the conjugate PF$_{5}$ is a Lewis acid and
can form adducts with GO.
Also note that similar advantages          
have been seen on the performance of Li-ion batteries
with Lewis adducts of BF$_{3}$ applied as electrolyte additives 
\cite{nie2016some,xiao2016lewisadducts,matsui2014,wang2000,carino2015bf}.

All-solid-state Li and Na ion batteries have also been built and tested where GO plays the
role of cathode active species, solid electrolyte and separator at the same time \cite{ye2017metal}. 
However, the capacity of this batteries quickly decreases upon cycling, likely for at least  
the lack of stabilizing BX$_{3}$ additives as seen for the above liquid electrolyte GO batteries.

\subsection{Basal plane functionalization in h-BN cathodes and elsewhere}
Another recent example of rechargeable 2D cathode actives species is based on h-BN
functionalized with fluorene on its edges and defect sites \cite{venkateswarlu2019effective}. 
Contrary to the expectation
that a fluorinated h-BN would behave similarly to graphite 
fluoride and form a non-rechargeable
battery \cite{CFxBattery1970}, Ref. \onlinecite{venkateswarlu2019effective} demonstrates
rechargeability of F-doped h-BN cathode with a Mg anode at 50 mAh/g capacity over 80 cycles. 
Also note that it is extremely difficult to make stable basal plane functionalized fluorinated
h-BN, and to date it was only possible by a high energy ion irradiation of a
LiF/h-BN/Cu heterostructure \cite{entani2020non}.

While the reversible capacity of GO cathodes is largely associated with   
edge and defect functionalization, the utilization of the basal plane
functionalization for reversible redox processes is still to be developed
and is very desirable. A close packing of the basal plane with redox active
functional groups can potentially enable large capacity without the
generation of edges and defect sites in the 2D material. Also note the thermal instability of 
GO that is the main hurdle for its industrial scale production and its limited safety in
applications
\cite{kim2010self,krishnan2012energetic,lowe2017challenges,lin2019synthesis}.

Free radicals appear to be the best and most general means for a
stable covalent functionalization of the basal plane. Graphite fluoride or graphite
oxide can both be considered as functionalized by fluorene or oxygen
radicals. Radical functionalization of the basal plane of h-BN has been
carried out with $\cdot$OSO$_{2}$F \cite{bartlett1978novel,shen1999intercalation}, 
$\cdot$NH$_{2}$ \cite{ikuno2007amine,nazarov2012functionalization,RU2478077C2}, 
$\cdot$O$\cdot$ and ozone 
\cite{pakdel2014plasma,cui2014large,li2014strong,anota2011lda,nemeth2017OzoneBF3}, 
$\cdot$OH \cite{nazarov2012functionalization,RU2478077C2,sainsbury2012oxygen},
$\cdot$SO$_{3}$H \cite{RU2478077C2,mofakhami2010material,mofakhami2011material,mofakhami2015material}, $\cdot$R \cite{rajendran2012surface,sheng2019polymer} 
and $\cdot$OR \cite{sainsbury2012oxygen} (R being alkyl or aryl),
and :CBr$_{2}$ \cite{sainsbury2014dibromocarbene} radicals. 
These radicals preferentially bind to the basal plane B atoms
of h-BN as its valence shell is incomplete and can in principle accept two
more electrons \cite{nemeth2018simultaneous}. 

Lewis adduct formation
between the B of h-BN and the N of the amino groups of non-radical amines has also been reported
\cite{lin2009soluble}, however it has not been conclusively established that
the formation of Lewis adducts happens at the edges/defects or on the basal
plane. Such Lewis
adducts should preferentially form between edge or defect B atoms and amino
groups as the edge atoms of h-BN clearly have strong Lewis acid/base
character. Basal plane atoms of h-BN are much more difficult to bring into a
non-planar hybridization necessary for Lewis adduct formation, albeit folding
and bending of the h-BN sheet may increase Lewis acid/base character of
some surface atoms.

Hydroxide anions covalently functionalize the edges or defect sites of h-BN, but 
not the basal plane, when
h-BN is ball-milled and exfoliated in aqueous NaOH solution \cite{lee2015scalable}. 
Edge functionalization of h-BN is likely dominant also in the melt phase of hydroxides 
\cite{zhang2016experimental}. The reaction of h-BN with strong Lewis base nitrides, such as 
Li$_{3}$N, is assumed to go through the formation of a complex with
approximate composition Li$_{3}$N.3BN before it results in Li$_{3}$BN$_{2}$ and cubic BN 
\cite{wentorf1961synthesis}. This indicates that the reaction of h-BN with very 
strong Lewis bases may result in a basal plane functionalized h-BN, however these complexes likely 
have limited thermal stability. 

Recently discovered cyanographene and its derivative graphene acid are also examples of basal
plane functionalized 2D materials and have been successfully tested for
supercapacitor applications 
\cite{bakandritsos2017cyanographene,cheong2019cyanographene,talande2019densely}.
While the -CN and -COOH functionalizations of the basal plane of graphene
allow for the binding of redox active species, such as iron
oxides/oxyhydroxides \cite{talande2019densely} the -CN and -COO(H) groups themselves
do not appear to participate in redox reactions, hence these materials in
themselves can be applied only as electrodes in double layer capacitors,
although they provide great improvement in capacity 
as compared to non-functionalized
graphene \cite{cheong2019cyanographene}. Also, the density of
functionalization is only about 15\%, which is large for most known
functionalized graphenes, but still much smaller than in
fluorographene, graphene (mon)oxide, or in (BN)$_{2}$OSO$_{2}$F (made of highly oriented
pyrolytic h-BN) \cite{shen1999intercalation}. 

Nitrogen doped graphene also contains basal-plane bound nitrogen atoms and
greatly improves the capacity of graphene supercapacitors \cite{jeong2011nitrogen}.
The N atoms mostly dope the basal plane (substitute C atoms) instead
of functionalizing it and participate only to a limited extent in redox
reactions. All these latter systems (N-doped and -CN or -COO(H)
functionalized graphene) need an external electrolyte, because they cannot
provide solvation shells to cations due to the insufficient density of functional
groups or the in-plane nature of the doping. 

As an example of basal plane functionalized cathode active species,
Li[(BN)$_{2}$OBF$_{3}$] (discharged form) has been proposed in our previous work
\cite{nemeth2015functionalized,nemeth2018simultaneous,nemeth2020-US10693137}.
This material is closely related to the above
mentioned $\cdot$OSO$_{2}$F functionalized h-BN, however, instead of a
charge-neutral radical, an anionic radical, $\cdot$OBF$_{3}^{-}$ 
functionalizes the B atoms of
h-BN. The radical character of the anion ensures its covalent binding to
h-BN while its negative charge enables the holding of cations between the
$\cdot$OBF$_{3}^{-}$ groups on the surface of h-BN. Furthermore, when densely
functionalized, this system also enables the solvation of cations
without external electrolyte and presents a solid ion conductor.

In addition to excellent ionic conductivity, the electronic conductivity of 
Li[(BN)$_{2}$OBF$_{3}$] is also predicted to be great, due to a high concentration of 
holes in the N lone pairs (p$_{z}$ orbitals). These N-holes are generated
by the strongly electron withdrawing character of the $\cdot$OBF$_{3}^{-}$ radicals
that withdraw almost a whole electron charge from the two N-s per formula unit
\cite{nemeth2018simultaneous}. We have recently pointed out the existence and charge storage
ability of similar N-hole states in Li$_{3}$BN$_{2}$ cathode active species \cite{emani2019li3bn2}.
The electronic conductivity of 
Li[(BN)$_{2}$OBF$_{3}$] is expected to be close to that of (BN)$_{2}$OSO$_{2}$F, 
which was measured to be 1.5 S/cm \cite{shen1999intercalation}. This electronic
conductivity is in the same order of magnitude with graphene (7.14 S/cm)
\cite{sainsbury2016covalent,rani2010electrical}
and electroconductive carbon additives (2-12 S/cm) \cite{pantea2003electrical} 
typically used in batteries. The concentration of N-holes and thus the electronic conductivity
of Li$_{x}$[(BN)$_{2}$OBF$_{3}$] depends on its oxidation state: the largest concentration 
of holes is in the fully oxidized state (x=0) and it gradually decreases as this system becomes
fully reduced (x=2). The as synthesized state (proposed below) is at x=1 and corresponds to the
half reduced system.

Therefore, Li[(BN)$_{2}$OBF$_{3}$] (and the analogous Na and Mg
compounds) can potentially provide three functions of the cathode in a single material: ionic and electronic
conduction and electroactive species. The combination of these features in a single material can
therefore save a number of solid phases that would normally be necessary in a composite cathode
and eliminates interfacial impedance on the respective phase boundaries \cite{pervez2019interface}.

\subsection{Graphene vs h-BN for functionalization}
As described in Refs \onlinecite{bartlett1978novel} and 
\onlinecite{shen1999intercalation}, $\cdot$OSO$_{2}$F functionalization
of graphene has been carried out in the form of first stage intercalation complexes
of graphite of C$_{8.5}$OSO$_{2}$F stoichiometry. 
Functionalization of graphene with $\cdot$OBF$_{3}^{-}$ radical
anions is also possible, however the product is expected to be much less thermally
stable as volatile decomposition product carbon-monoxide is expected
(besides BF$_{3}$). 

The surface density of functional groups on
graphene is expected to be significantly lower than in h-BN. This is indicated
by the stoichiometry of C$_{8.5}$OSO$_{2}$F as well, as compared to 
(BN)$_{3}$OSO$_{2}$F or even (BN)$_{2}$OSO$_{2}$F reported in Ref
\onlinecite{shen1999intercalation}. Similar stoichiometry
of C$_{6.6}$CN has been seen for base-plane functionalized 
cyanographene and for its derivative
graphene acid as well \cite{bakandritsos2017cyanographene}, despite that the parent
fluorographene can have equal number of C and F atoms. While oxygen
functionalized graphene can have a 1:1 carbon-to-oxygen ratio in graphene
monoxide \cite{mattson2011evidence}, 
the epoxy and hydroxyl groups on the surface of graphene can
diffuse, agglomerate and react with each other leaving the surface of
graphene as O$_{2}$ or 
H$_{2}$O molecules at temperatures over 70 $^{o}$C \cite{zhou2013origin}.
Such a decomposition of graphene oxide is even faster in the presence of
certain salts, such as potassium salts \cite{kim2010self}.
No such fast thermal decomposition is known for oxidized or hydroxyl
functionalized h-BN. In fact, oxygen functionalized h-BN is known to be stable
over 800 $^{o}$C \cite{li2014strong}.

The hopping of radicals between nearby carbon atoms on the surface of graphene
is much less hindered energetically than in h-BN 
and leads to recombination of radicals
between neighboring sites and ultimately to the detachment of these
recombined radicals from graphene or even oxidation of carbon atoms.
Fluorographene is the only known functionalized graphene that 
preserves a high degree of functionalization in a broad thermal window, up
to 400 $^{o}$C, albeit some decomposition starts on it above 100 $^{o}$C
\cite{devilliers1983mass}.
In the base-plane functionalized h-BN, 
radicals are bound to boron atoms and the nitrogen atoms form
barriers between nearest neighbor borons for the hopping of radicals hindering
recombination reactions. For this reason, h-BN appears more suitable for
dense, thermally stable surface functionalizations than graphite.

\subsection{Edge vs basal plane functionalization}
Both edge and basal plane functionalizations of 2D materials may be used  
as electroactive species in energy storage applications. However,
the basal plane normally provides much larger number of functionalization 
sites than the edges, therefore the utilization of the basal plane 
functionalization is
more advantageous for high capacity charge storage. The basal
plane functionalization can also play the role of solvation shell for
intercalating cations, i.e. it can be used as a solid electrolyte. This is
much less the case for edge functionalizations, unless too many edges and
lattice defects are introduced. However, too many
lattice defects may reduce mechanical and structural stability.
Therefore, basal plane functionalization appears inevitable for an
efficient energy storage device based on 2D materials. 

Graphene oxide (GO) normally contains both edge and basal plane functionalization.
Oxo (=O) groups functionalize the edges through double bonds to single C atoms
while epoxy (-O-) groups functionalize the basal plane through two single
bonds to two nearest neighbor carbon atoms. In order to utilize
graphene oxide as cathode electroactive species, one has to remove the epoxy
groups by gentle heating, because they would irreversibly react with Li or
Na and form Li$_{2}$O or Na$_{2}$O, as opposed to oxo groups which can 
be reduced and oxidized reversibly, at least when the temperature is not too
high \cite{liu2014lithium,jang2011graphene,kim2014novel}, albeit, as noted above
the selection of suitable electrolytes is also critical for the cycling 
stability of the oxo groups and the conversion of the oxo groups to -OBF$_{3}$ groups
appears to be one way to achieve this stability.

\section{Comparison of electronic and geometric structures of $\cdot$OSO$_{2}$F, 
$\cdot$OBF$_{3}$ and $\cdot$OBC\lowercase{l}$_{3}$ 
functionalized \lowercase{h}-BN}         

The robust synthesis and thorough characterization of $\cdot$OSO$_{2}$F
functionalized h-BN, (BN)$_{2}$OSO$_{2}$F forty years ago \cite{bartlett1978novel} 
and its instability toward
reductive species as well as its great electronic conductivity 
\cite{shen1999intercalation} and the great ionic
conductivity of functionalized boron nitrides in general as reviewed above,
motivated our re-design of (BN)$_{2}$OSO$_{2}$F in Ref. \onlinecite{nemeth2018simultaneous}
with the goal of achieving
a functionalization that preserves the excellent ionic and electronic conductivity
while also providing structural stability against reductive species. 
While in Ref. \onlinecite{nemeth2018simultaneous} the electronic structures of
oxygen, fluorene and $\cdot$OBF$_{3}$ functionalized h-BN were compared, here we
provide the first ever report on the band structure of (BN)$_{2}$OSO$_{2}$F
(synthesized and characterized forty years ago) as well as that of $\cdot$OBCl$_{3}$
functionalized h-BN and compare them to $\cdot$OBF$_{3}$ functionalized h-BN, along
with the band structures of the related Li, Na and Mg reduced species.
The effects of functionalization on the electronic and structural parameters 
of these species will also be discussed.

\subsection{Computational methodology}
Model systems for functionalized h-BN monolayers were built using a double stoichiometric
formula unit of the respective compounds in order to allow for
uniform functionalization on both the top and bottom surfaces of h-BN. Monolayers were
separated from each other by a 30 {\AA} distance that included about 22 {\AA} vacuum
layer while 3D periodic boundary conditions were applied.
Similar surface slabs are often used to model surface chemistry and physics 
\cite{GPacchioni05}.
Supercells of the simulation cells are shown in Figs. \ref{Structures} and \ref{StructuresBCl3}.

Density Functional Theory (DFT) calculations were carried out 
using the Quantum Espresso program package 
\cite{QE-2009,QE-2017}, following the methodology in 
our former works 
\cite{nemeth2014materials,nemeth2014ultrahigh,zhang2016experimental,nemeth2018simultaneous}. 
A plane wave basis set with 50 Ry
wave-function cut-off was used in conjunction with 
the PBEsol exchange-correlation functional
\cite{PBE,PBEsol} and the related ultrasoft pseudopotentials as provided by the software
package.
A k-space grid of the size of 10x10x4 (and twice this dense for h-BN) was employed to achieve mRy
convergence of the electronic energy with respect to k-space saturation.
This accuracy is satisfactory for the prediction of accurate cell
voltages within the chosen exchange-correlation functional and surface model. 
As the radical functionalization of h-BN often results in open shell systems,
spin polarized calculations were carried out.
Since the model systems have the same translational symmetry as
h-BN, only with larger unit cells, the representation of the electronic
band structure employed the same high symmetry k-points.
The coordinates of the high symmetry points in the k-space are $\Gamma$(0,0,0),
M(1/2,0,0) and K(2/3,1/3,0). 

The model systems were relaxed until residual forces on the atoms became smaller
than 0.0001 Ry/bohr and residual pressure on the simulation cell was less than 3
kbar. Energies of cell reactions were calculated from electronic energy differences
of products and starting materials, all in the crystal phase. Open cell voltages are 
computed as the negative of the cell reaction energies divided by the number of electrons
transferred in the reaction. Capacity density values refer to the total charge of
positive ions per mass of the discharged active material, expressed in units of
mAh/g. Energy densities are obtained as the product of the open cell voltage and
the corresponding capacity and expressed in Wh/kg units. 

L{\"o}wdin charges were calculated by projecting the electron
density from plane wave basis into atomic orbital basis. As the two representations span
different Hilbert spaces, the projection does not preserve charge neutrality 
for the sum of charges in atomic orbital basis.

Validation of this methodology can be found in our earlier works 
\cite{nemeth2014ultrahigh,nemeth2018simultaneous}
and it includes experimental
lattice parameters and enthalpies of formation of $\alpha$-Li$_{3}$BN$_{2}$, 
Li$_{3}$N, and h-BN. 
Experimental lattice parameters have been
reproduced within 2.5\% error while experimental enthalpies
within 4\% \cite{nemeth2014ultrahigh}.

The PBEsol functional, as a GGA type functional, typically underestimates band gaps,
for example by about 1 eV for h-BN \cite{zolyomi2015towards,nemeth2018simultaneous}.
Despite this inaccuracy, it is in general capable to correctly estimate the relative
size of band gaps in a group of similar compounds, such as functionalized h-BN
species. Higher accuracy methods, such as the PBE0
\cite{PBE01perdew1996rationale,PBE02adamo1999toward} or HSE \cite{heyd2003hybrid}
functionals are much more computationally expensive in a plane-wave basis therefore
PBEsol remains a practical tool for investigating and selecting designer molecules for
experimental testing.

\subsection{Effect of functionalization on properties}

\begin{figure*}[tb!]
\resizebox*{6.4in}{!}{\includegraphics{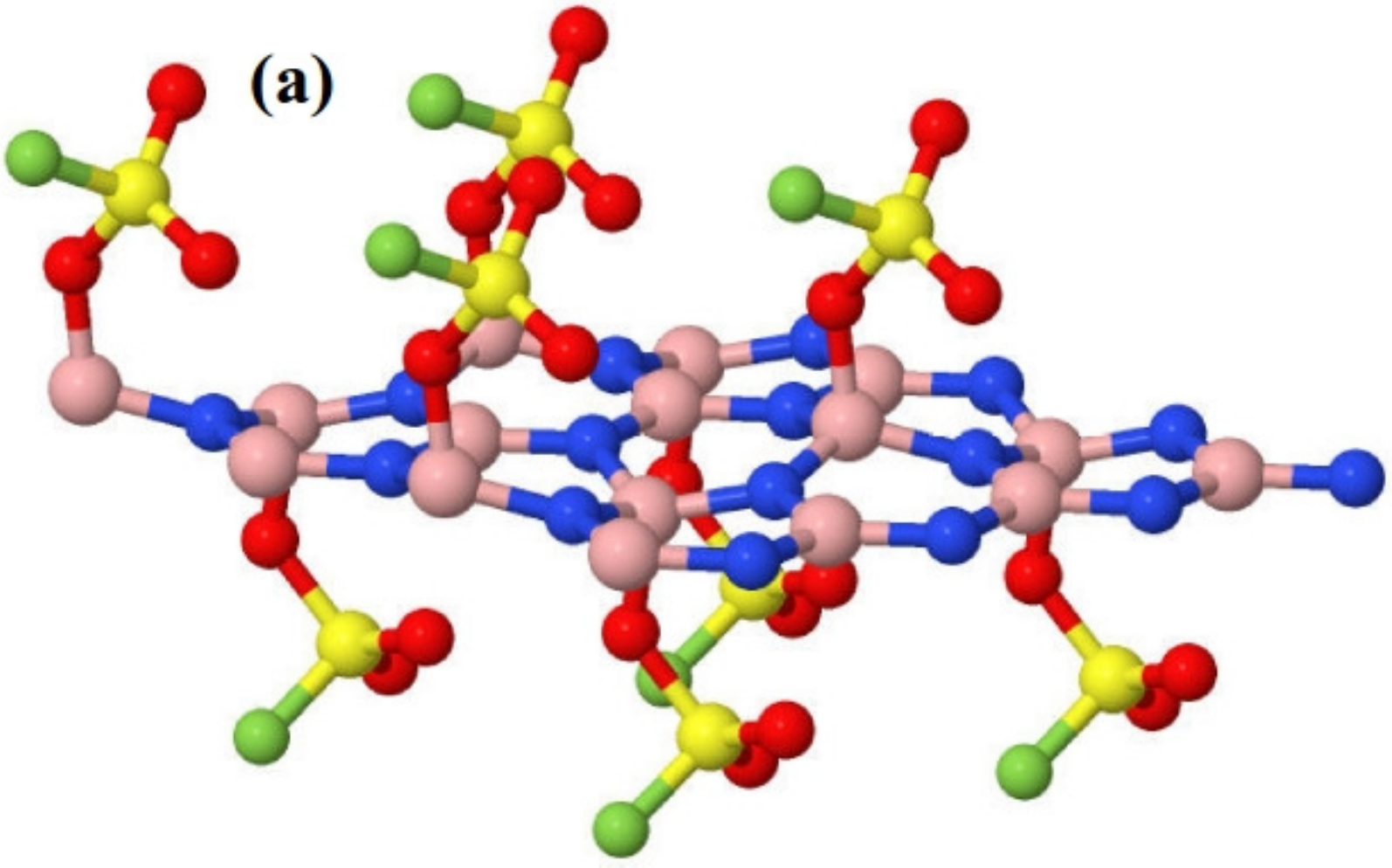}\includegraphics{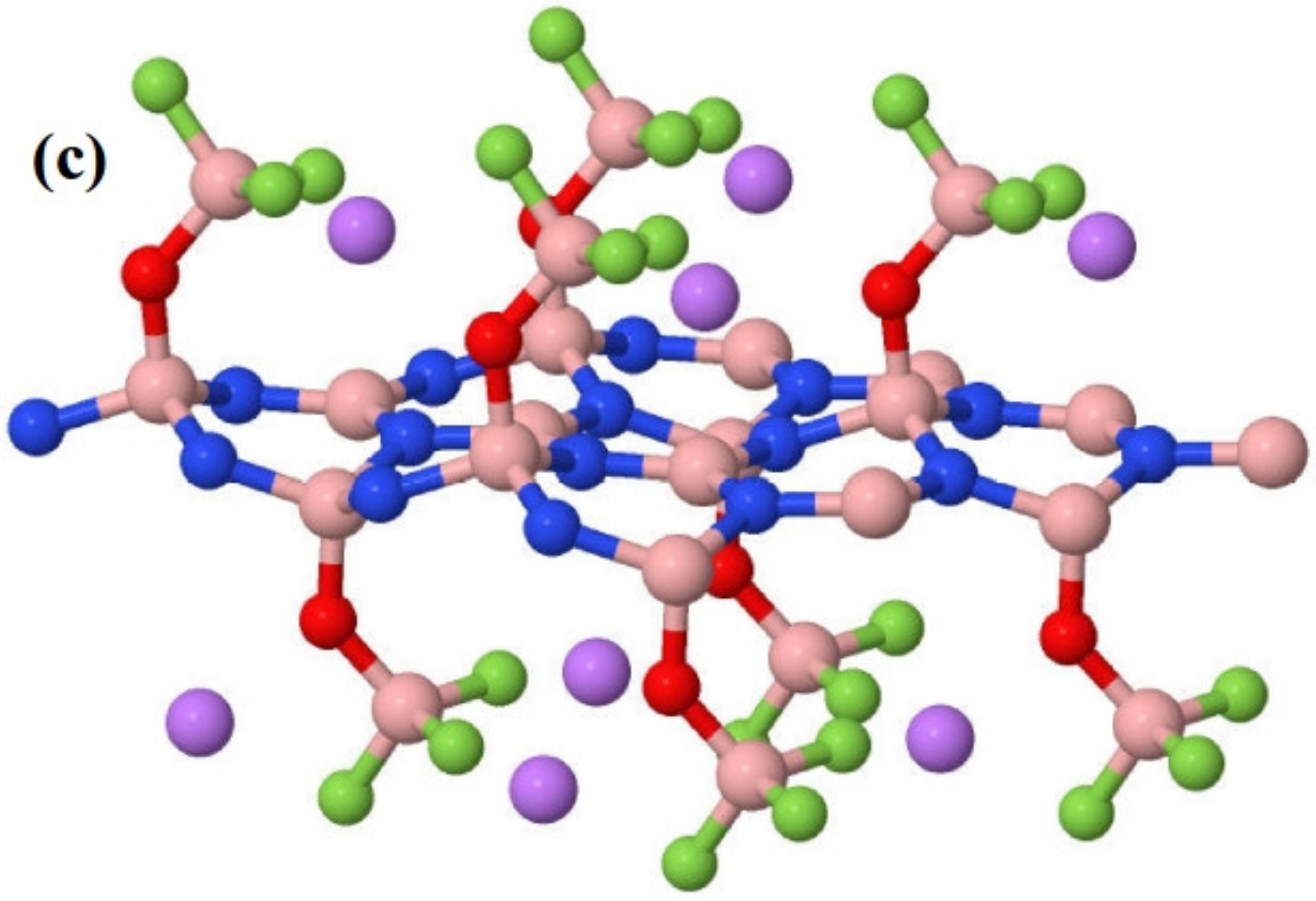}}
\resizebox*{6.4in}{!}{\includegraphics{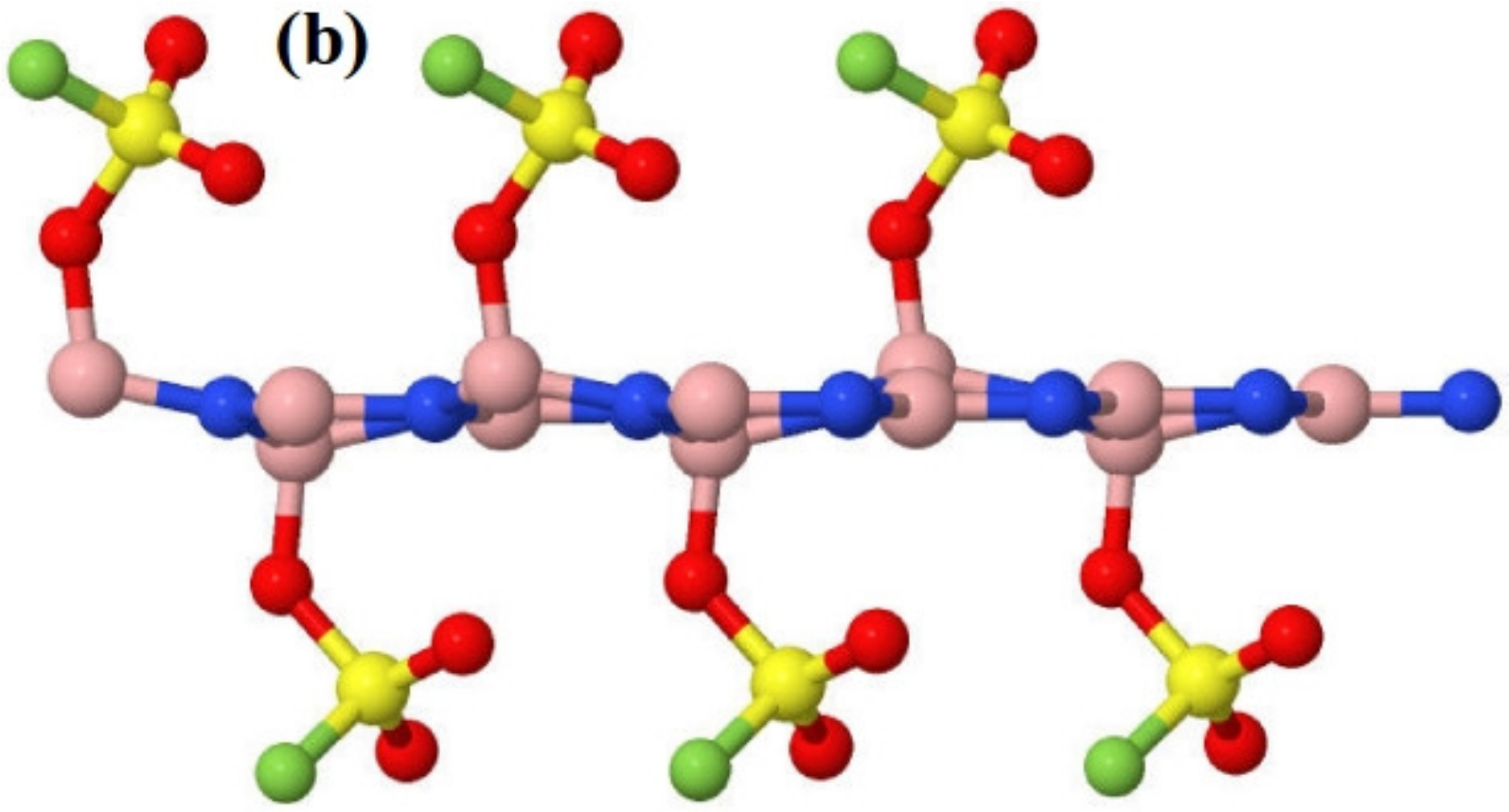}\includegraphics{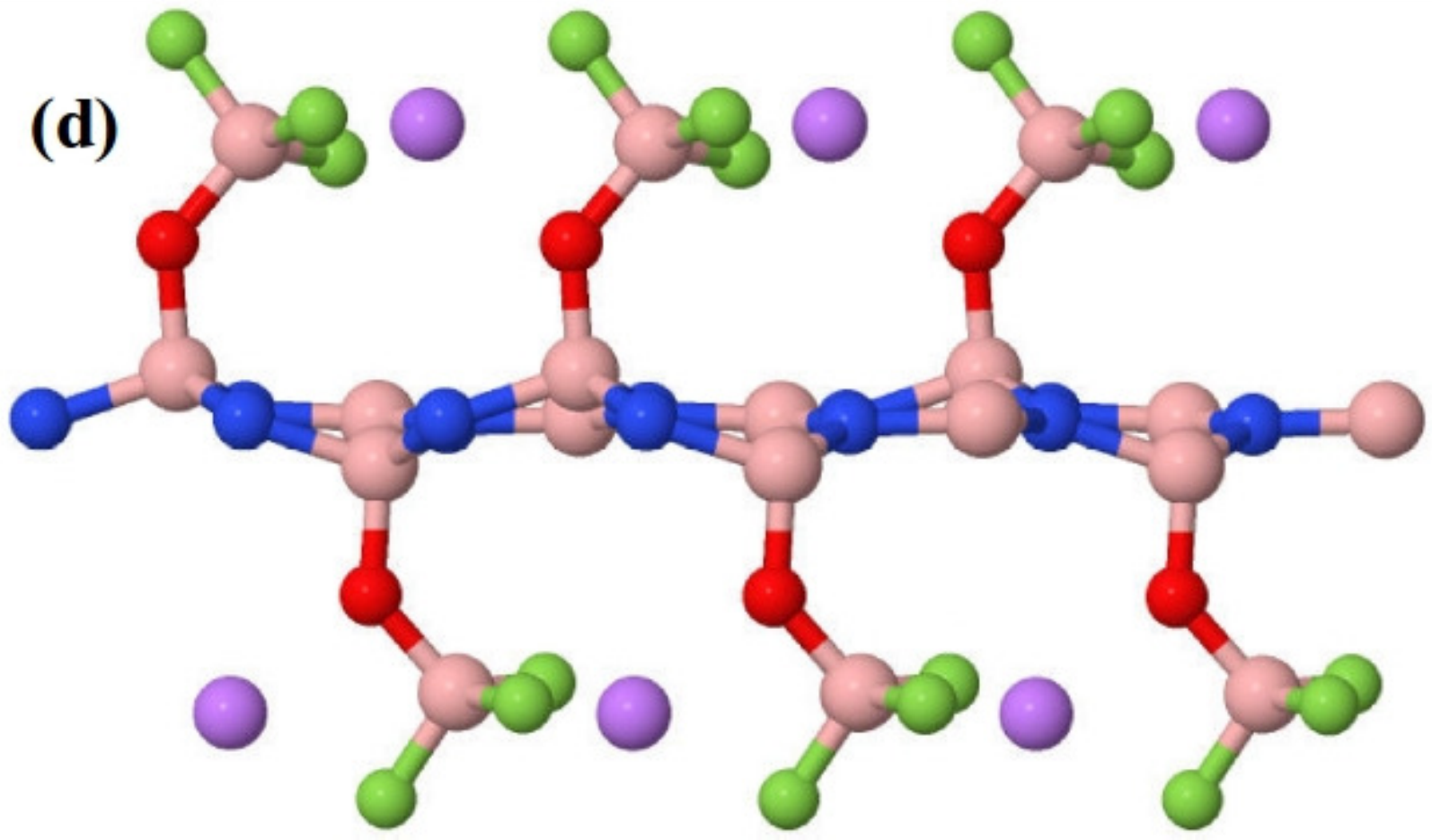}}
\caption{
Perspective and side views of 2x2 supercells of monolayers of
(BN)$_{2}$OSO$_{2}$F (panels a and b) 
and Li[(BN)$_{2}$OBF$_{3}$] (panels c and d). They represent maximum density
space filling functionalizations of the surface of a h-BN monolayer
with the $\cdot$OSO$_{2}$F radical and with
the $\cdot$OBF$_{3}^{-}$ radical anion, with associated Li$^{+}$ for the latter,
respectively. Color code: N -- blue, B -- bronze, O -- red, S -- yellow, F -- green,
Li -- magenta. Structures represent optimum energy geometries of the respective surface slabs.
}
\label{Structures}
\end{figure*}

\begin{figure*}[tb!]
\resizebox*{6.4in}{!}{\includegraphics{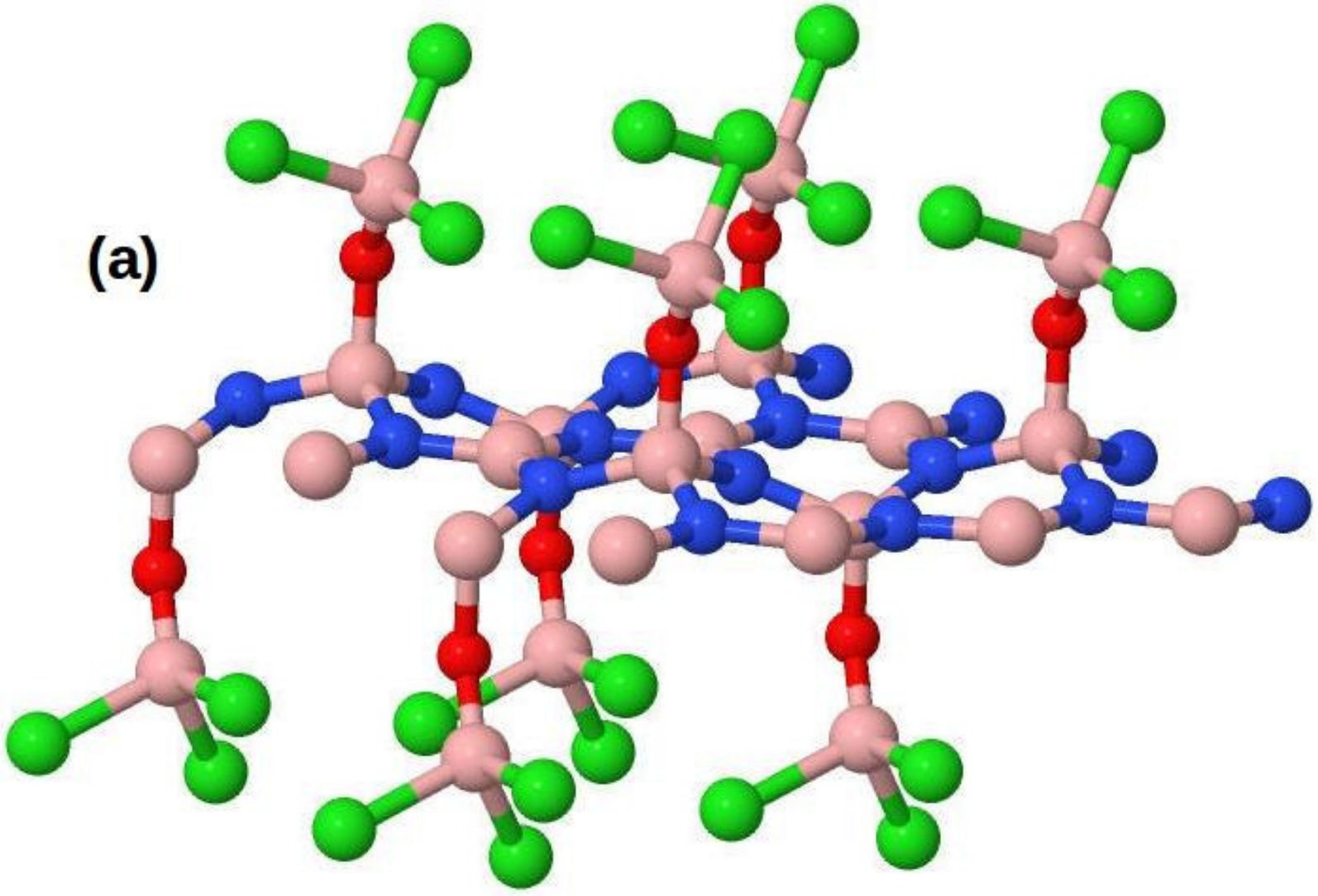}\includegraphics{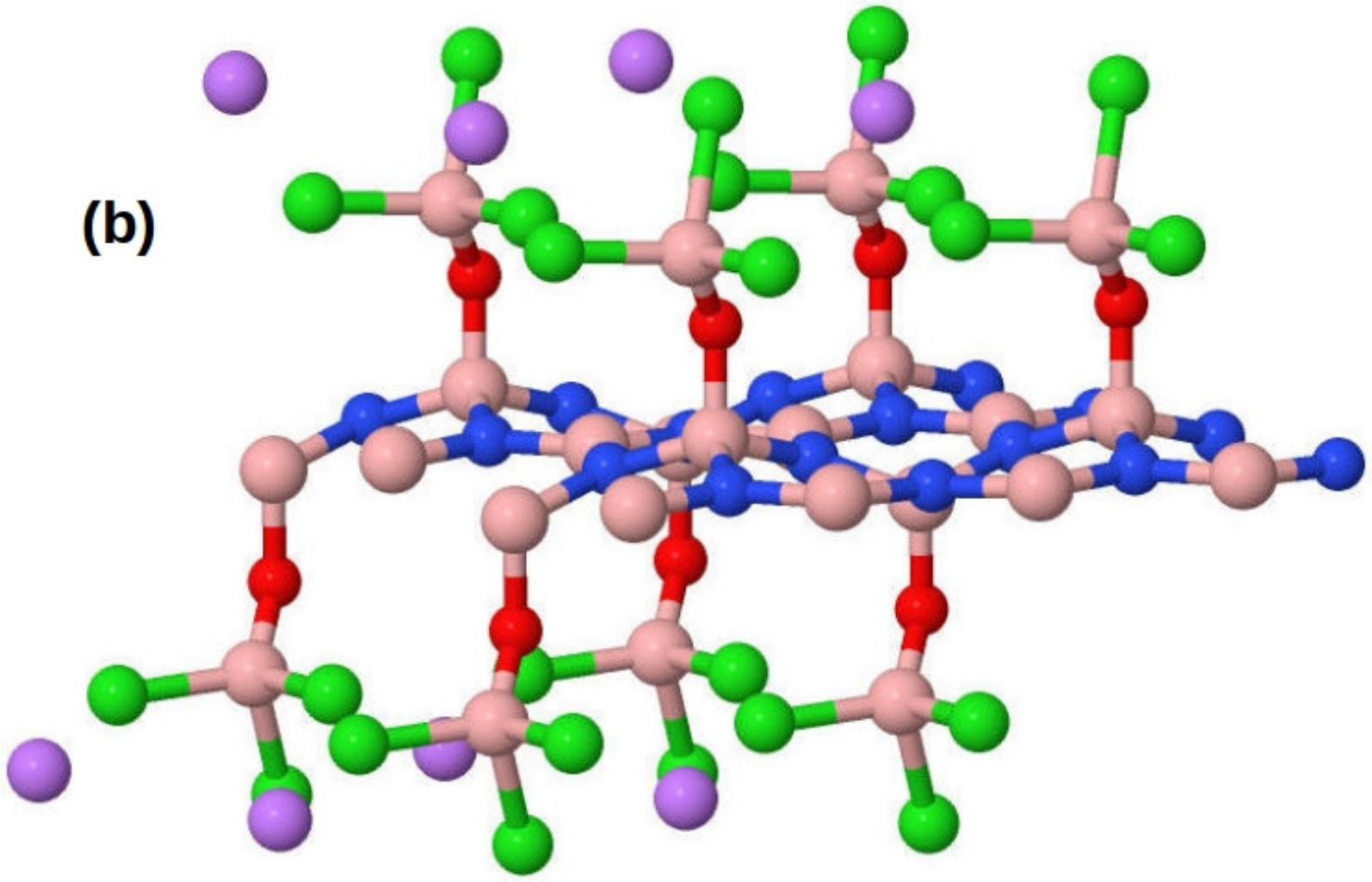}}
\resizebox*{6.4in}{!}{\includegraphics{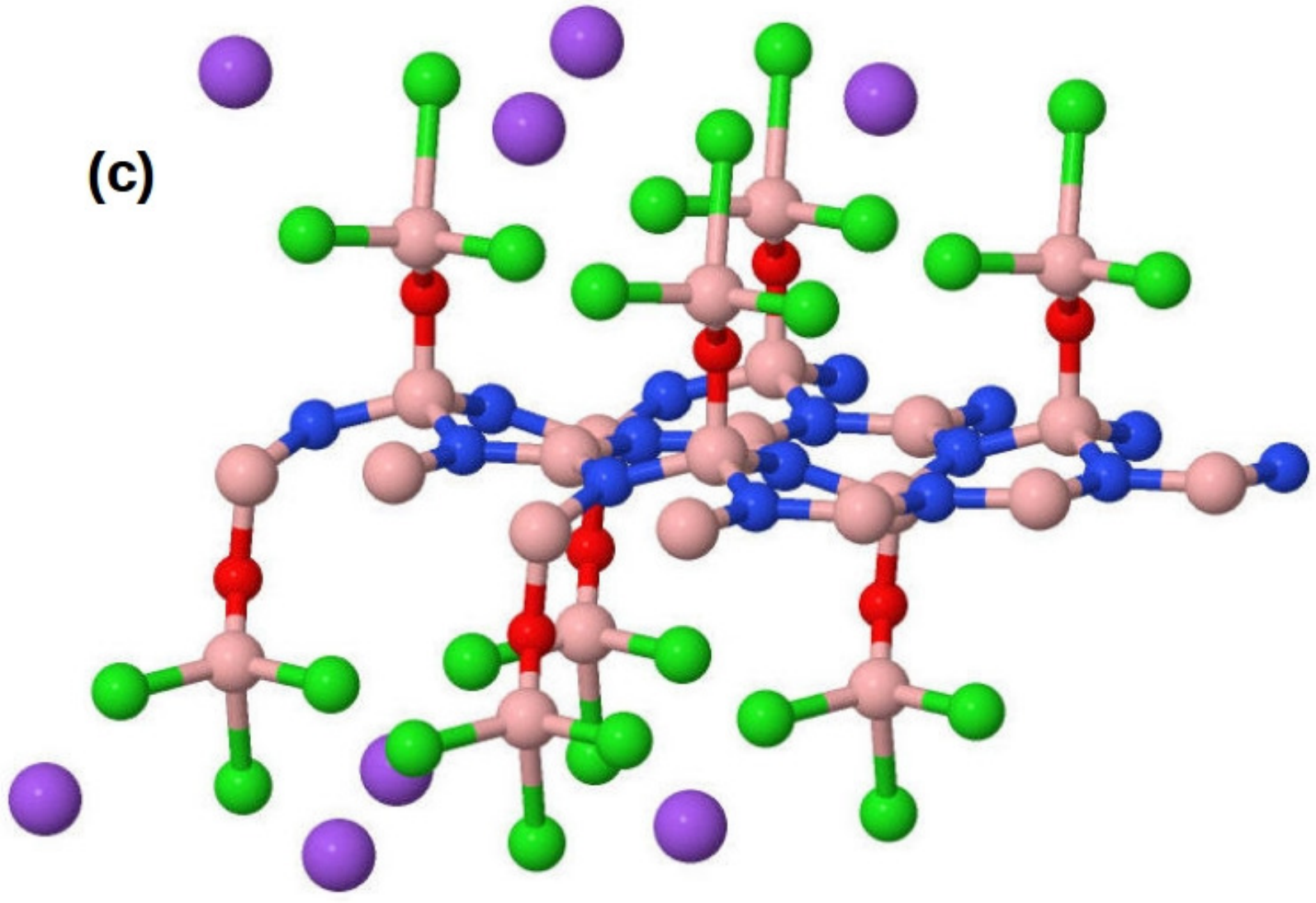}\includegraphics{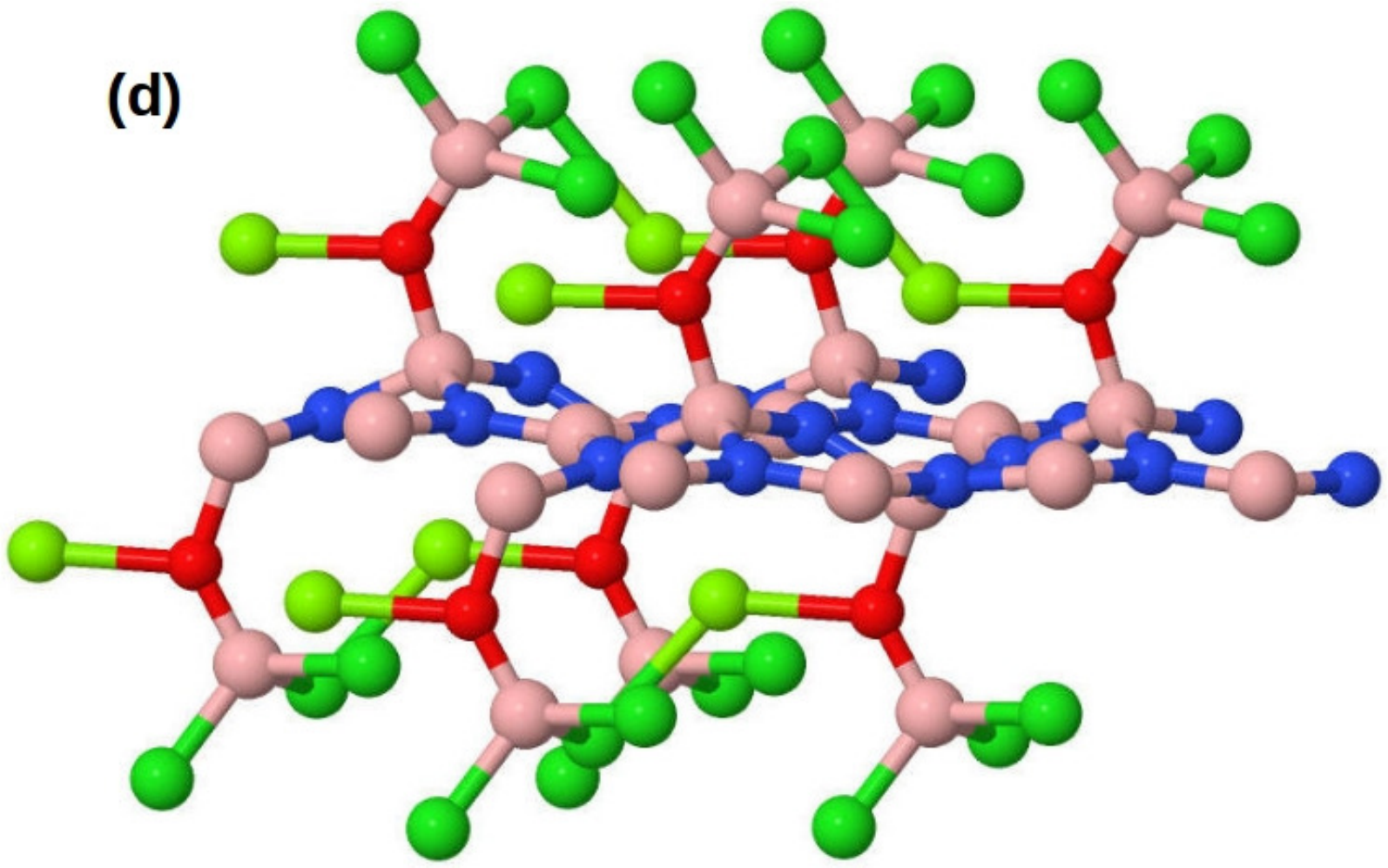}}
\caption{
Perspective views of 2x2 supercells of monolayers of 
(BN)$_{2}$OBCl$_{3}$ (panel a) 
and its Li, Na and Mg reduced versions with one reducing atom per formula unit:
Li[(BN)$_{2}$OBCl$_{3}$], Na[(BN)$_{2}$OBCl$_{3}$] and Mg[(BN)$_{2}$OBCl$_{3}$]  (panels b through d,
respectively). They represent maximum density
space filling functionalizations of the surface of a h-BN monolayer with the given species.
Color code: N -- blue, B -- bronze, O -- red, Cl -- green,
Li -- magenta, Na -- purple, Mg light green. 
Structures represent optimum energy geometries of the respective surface slabs.
Related structures with F instead of Cl in the OBCl$_{3}$ units have been depicted in our
former work 
\cite{nemeth2018simultaneous}.
}
\label{StructuresBCl3}
\end{figure*}

\begin{figure*}[tb!]
\resizebox*{6.4in}{!}{\includegraphics{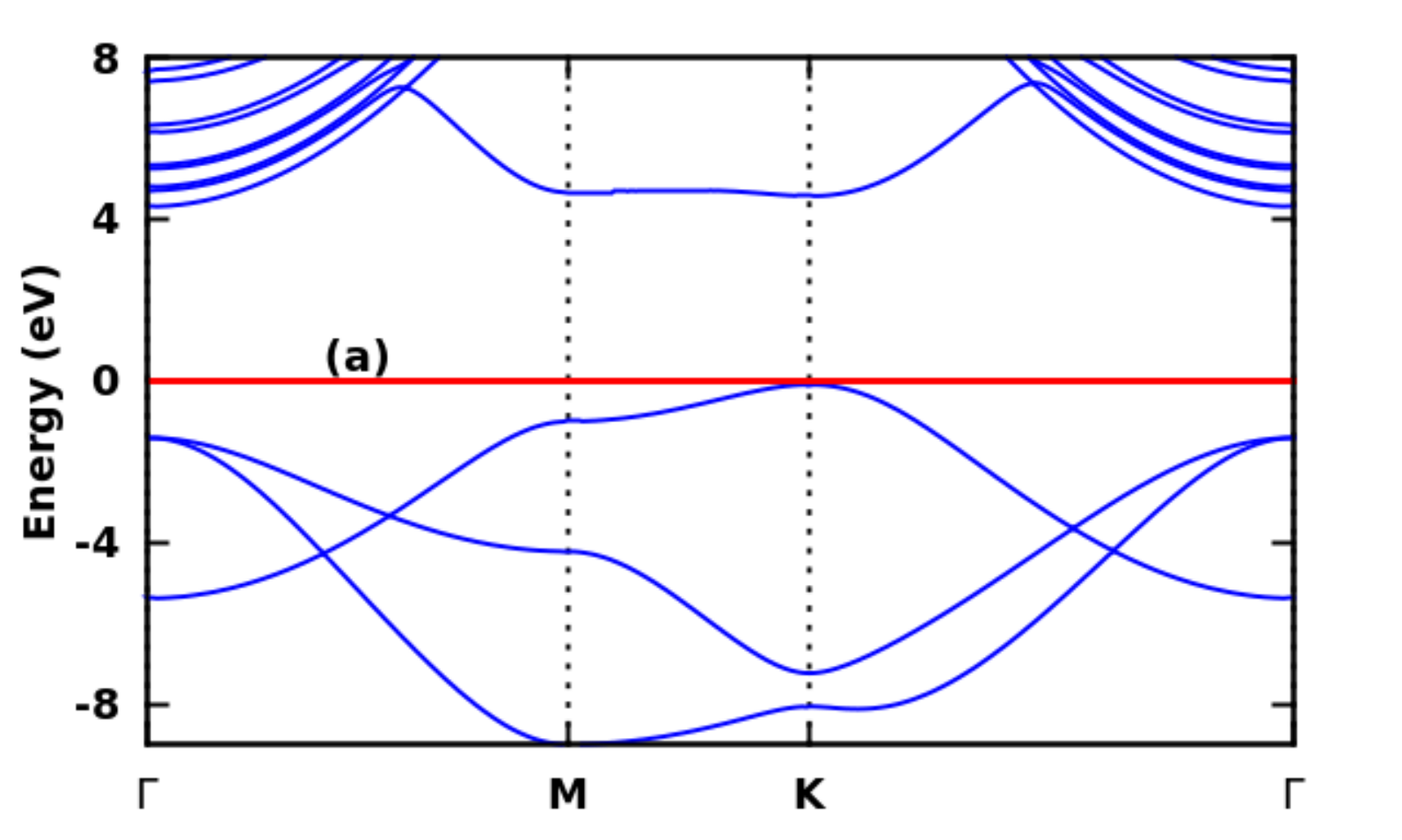}\includegraphics{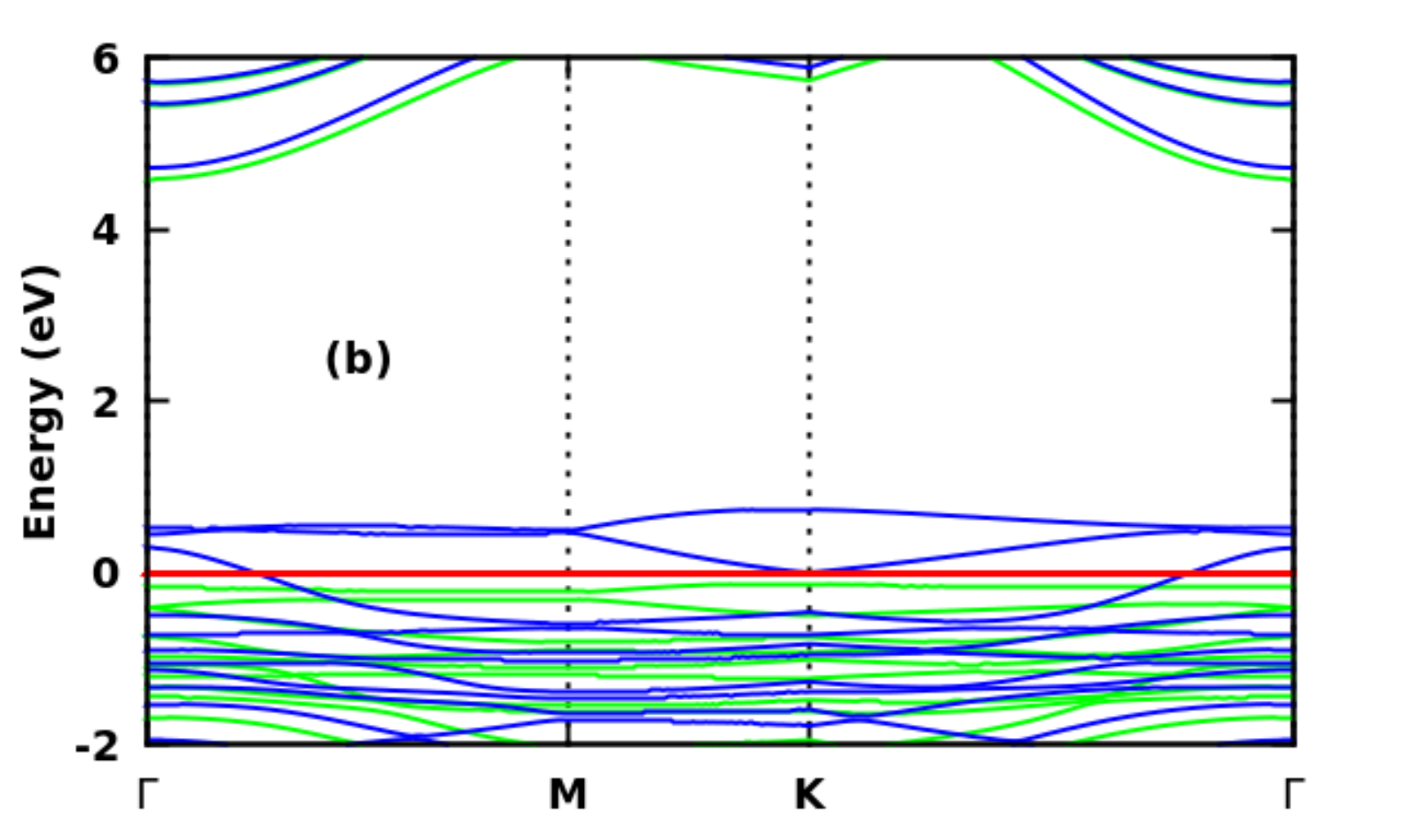}}
\resizebox*{6.4in}{!}{\includegraphics{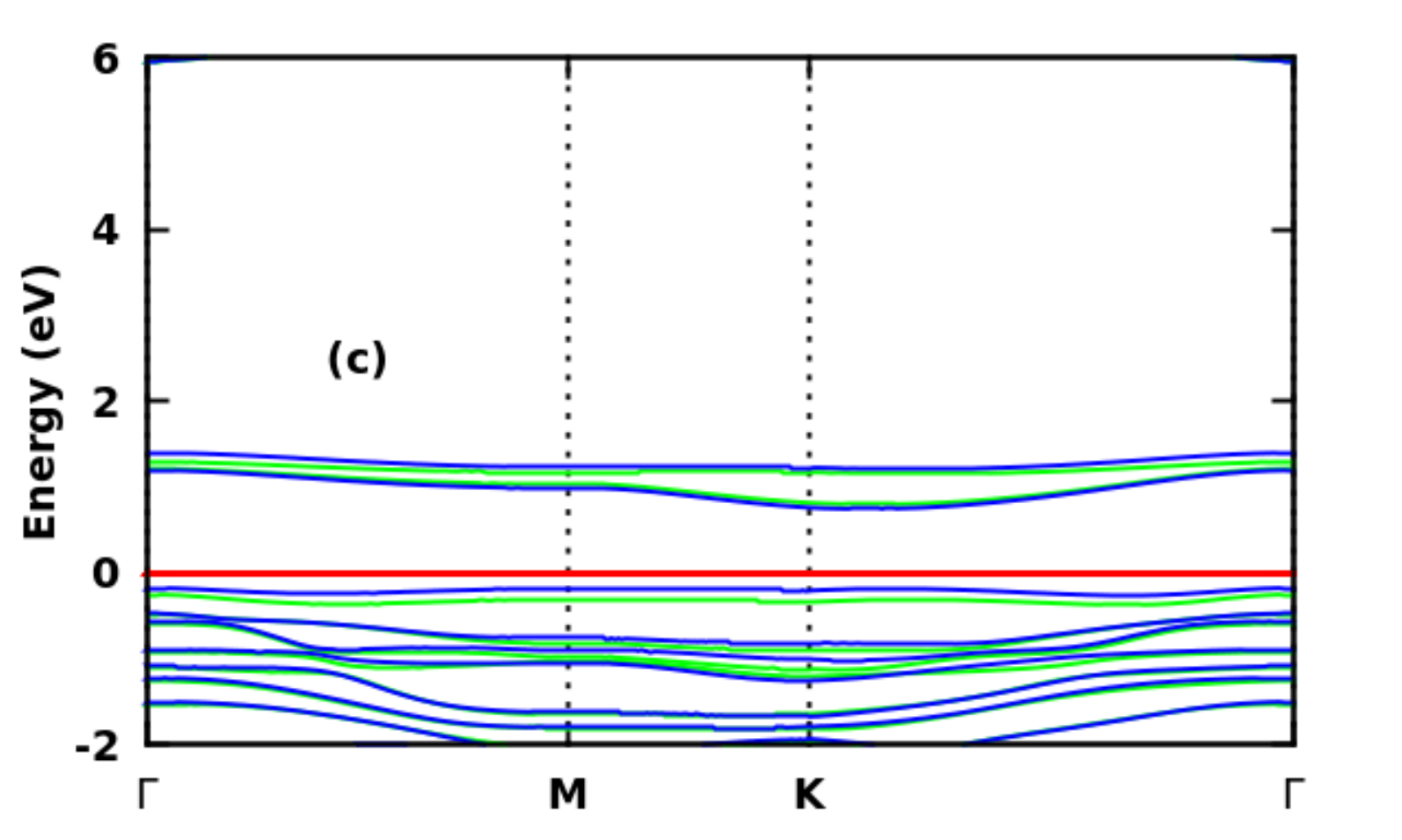}\includegraphics{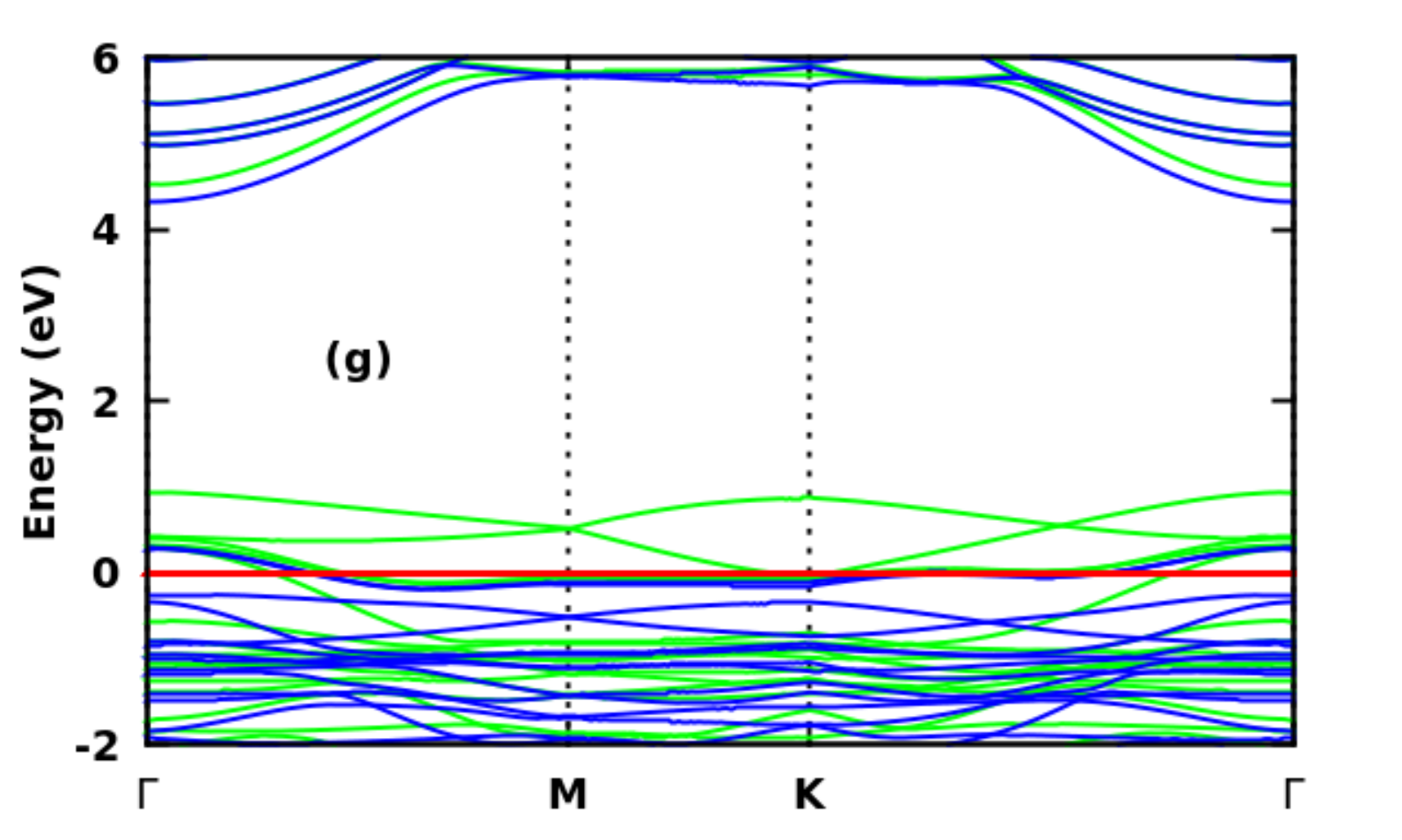}}
\resizebox*{6.4in}{!}{\includegraphics{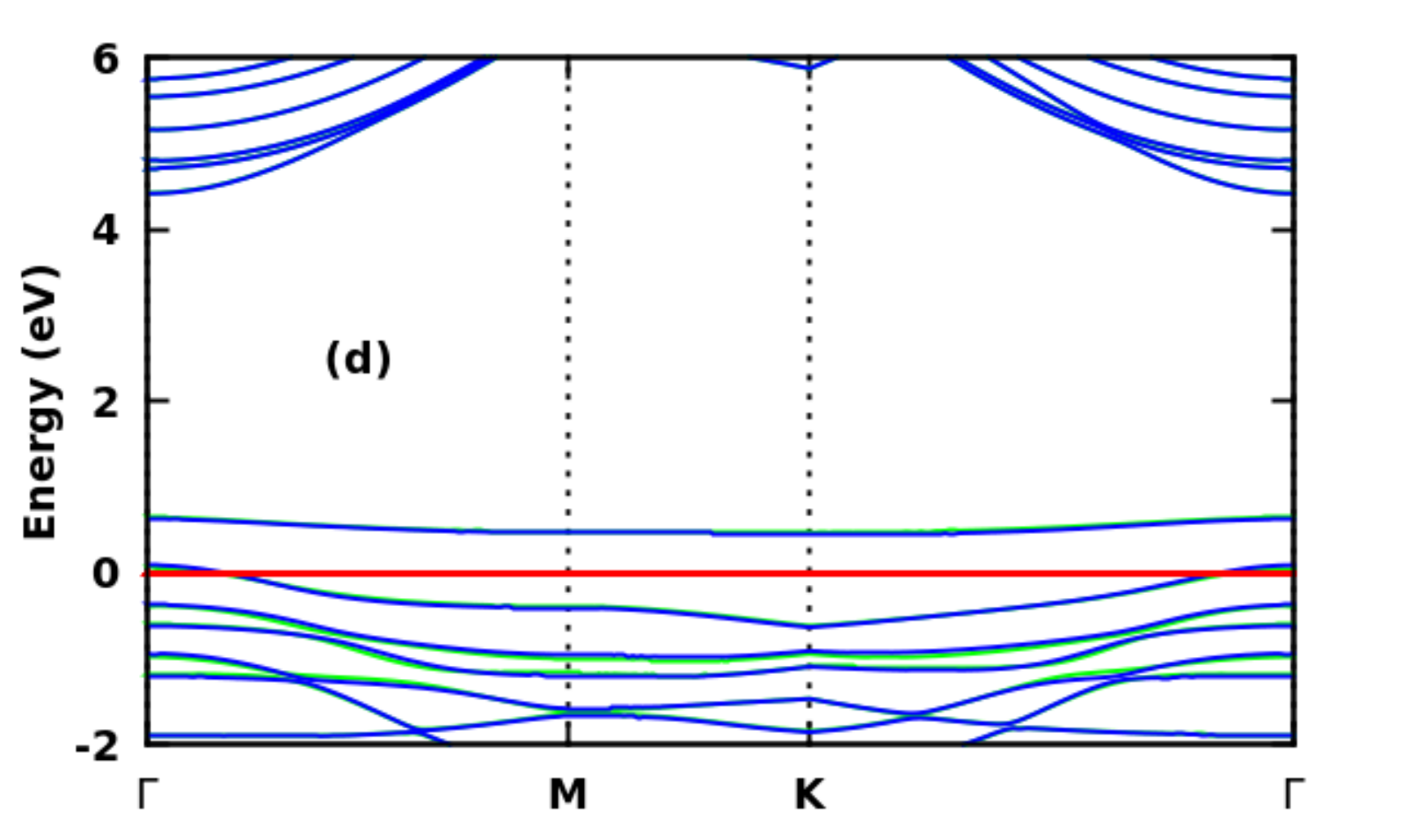}\includegraphics{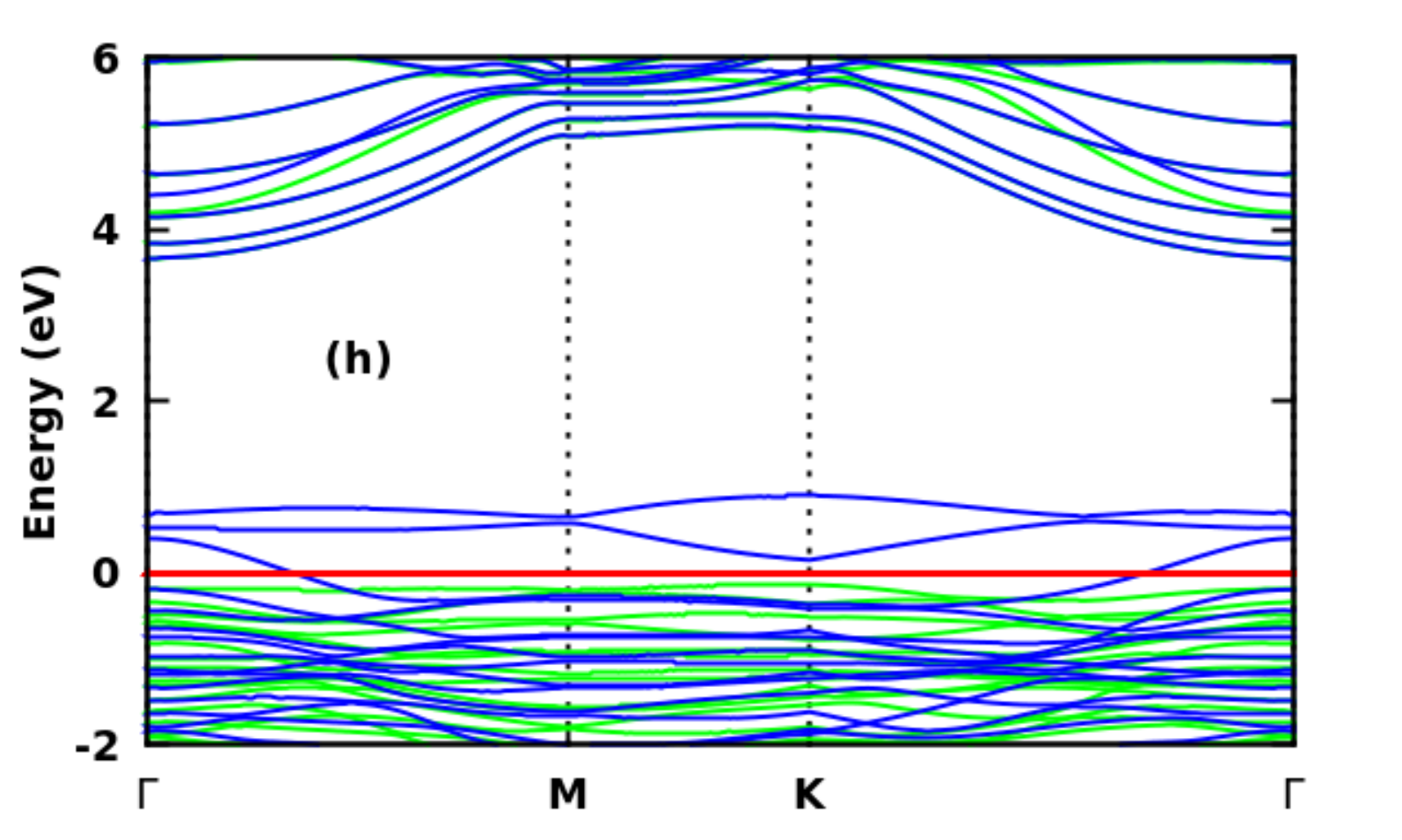}}
\resizebox*{6.4in}{!}{\includegraphics{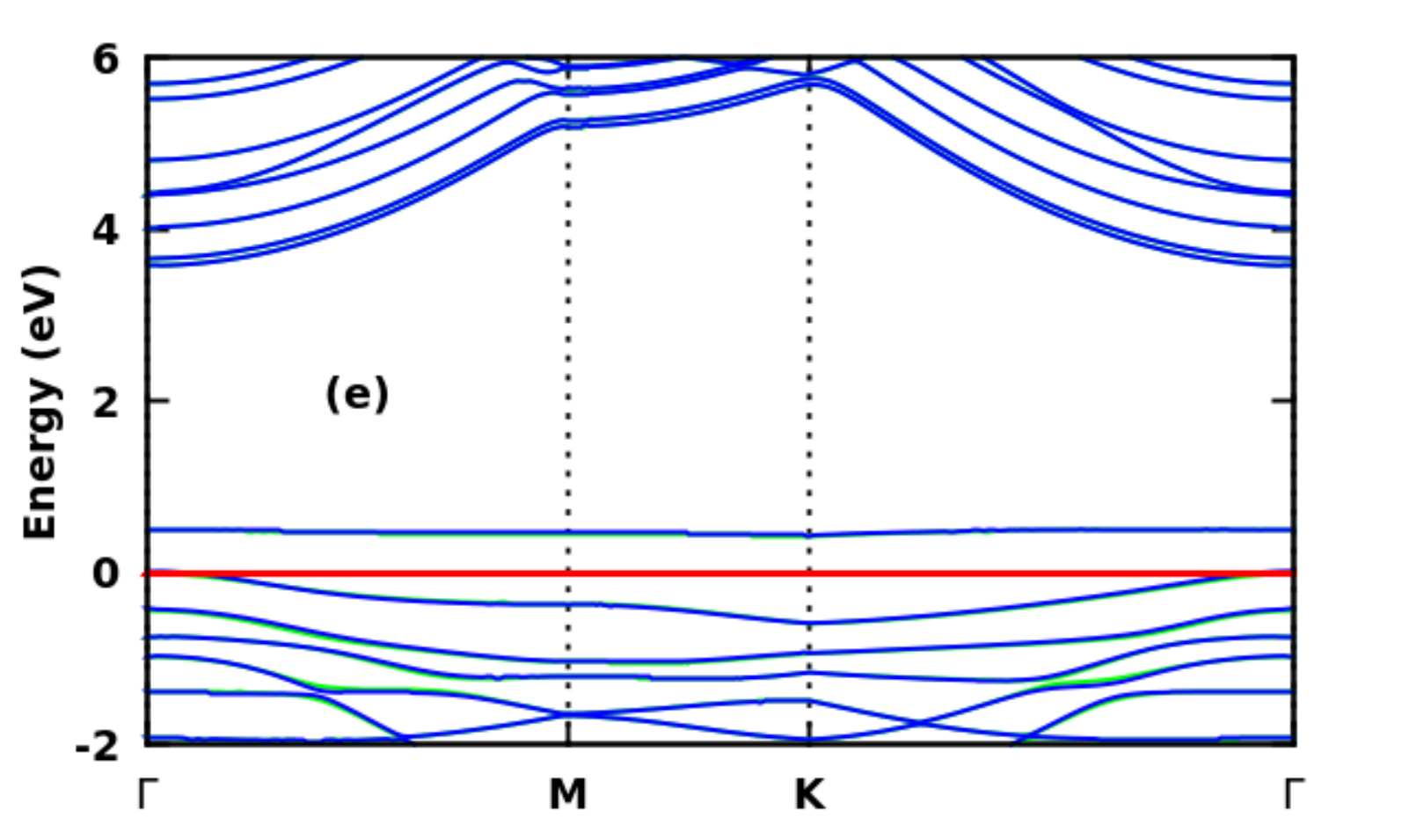}\includegraphics{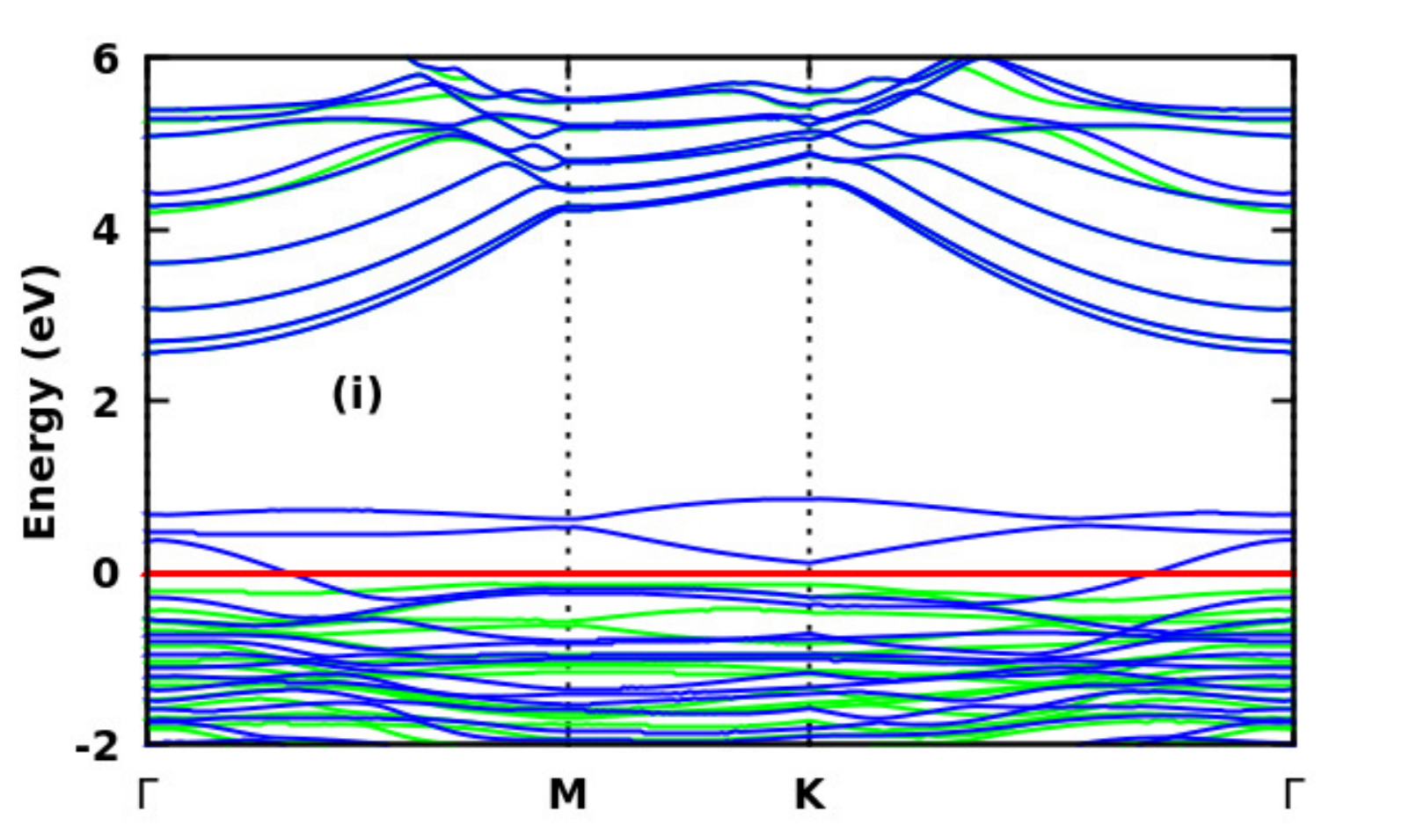}}
\resizebox*{6.4in}{!}{\includegraphics{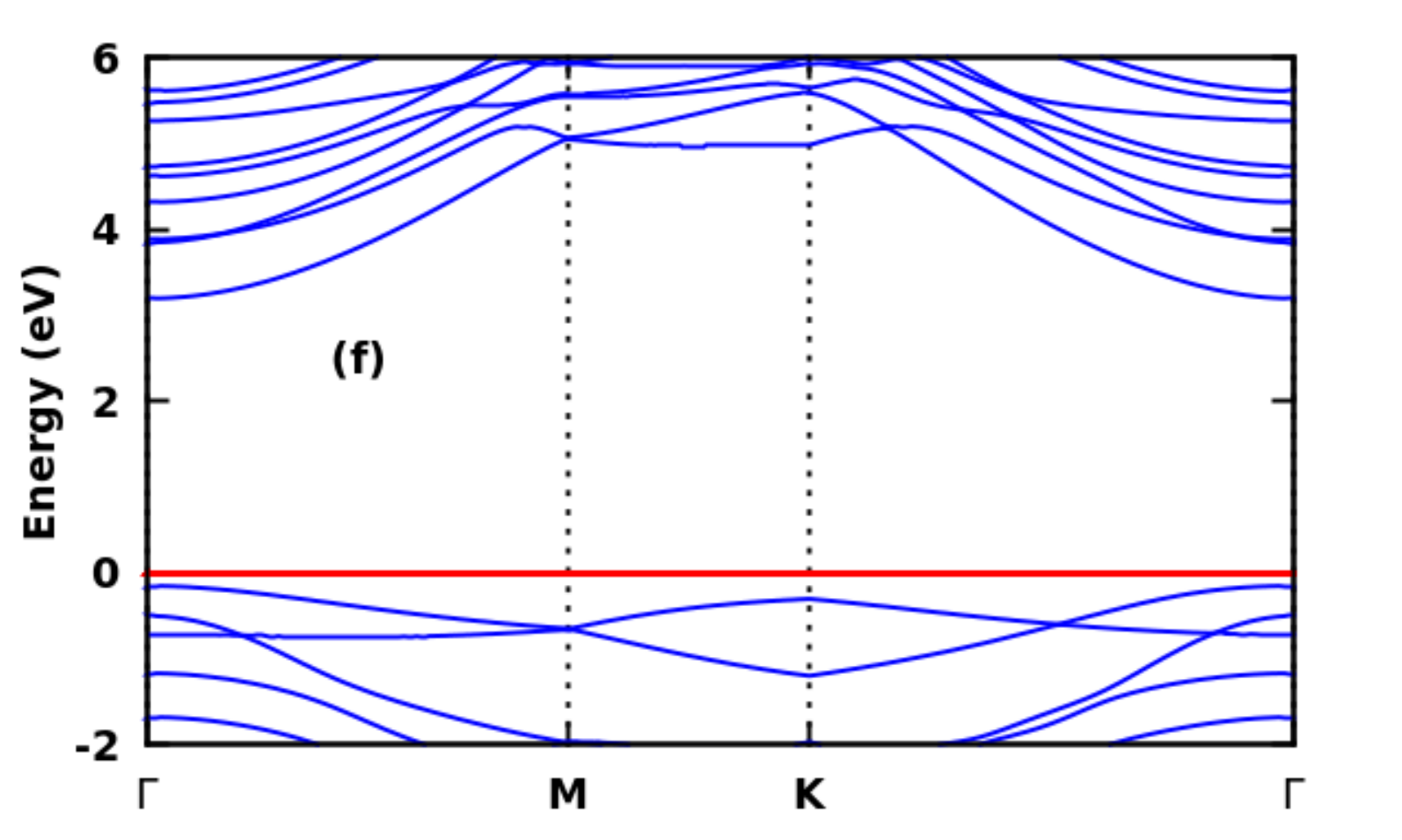}\includegraphics{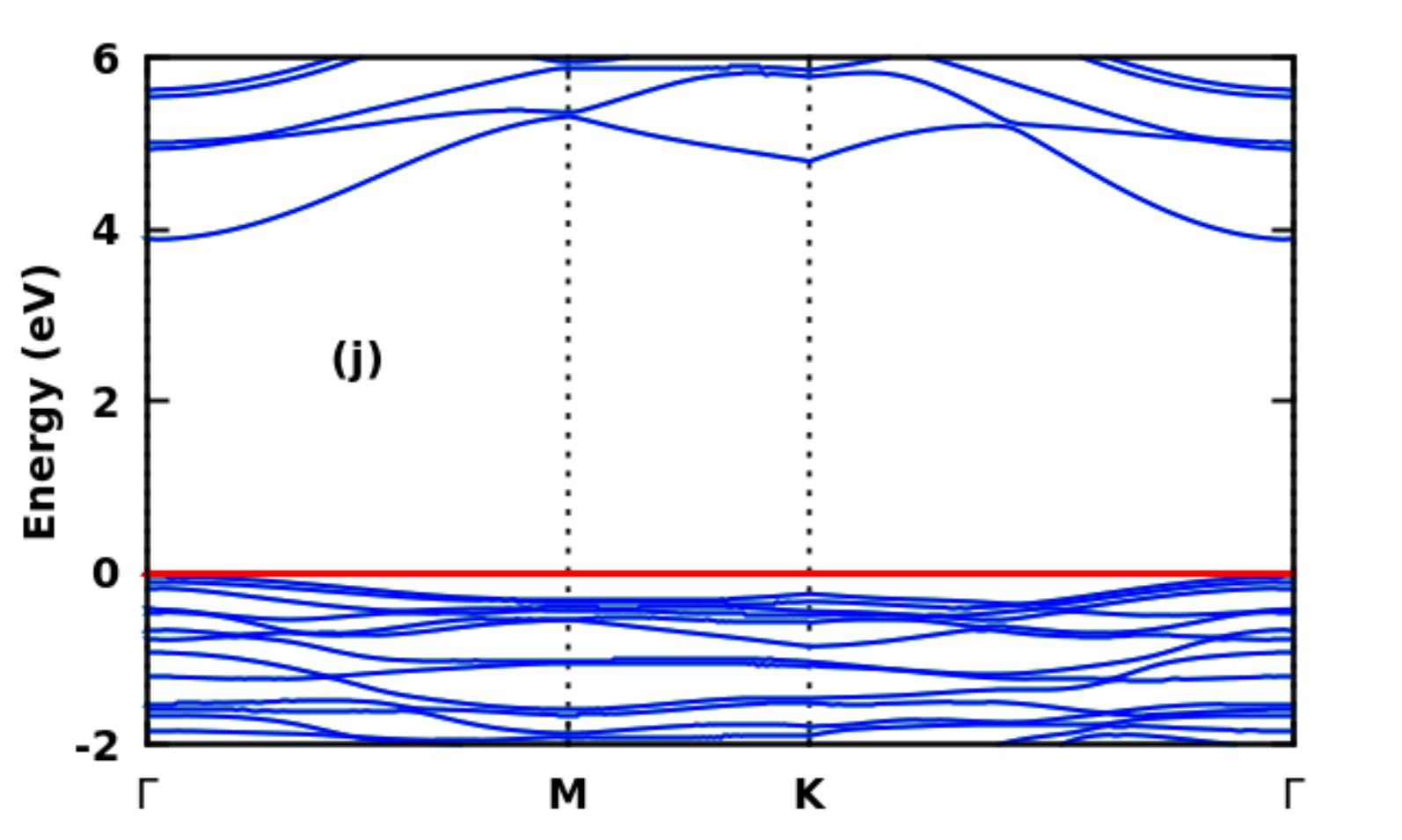}}
\caption{
Band structures of h-BN (a), (BN)$_{2}$OSO$_{2}$F (b), 
(BN)$_{2}$OBF$_{3}$ and (Li/Na/Mg)[(BN)$_{2}$OBF$_{3}$] (c-f), and
(BN)$_{2}$OBCl$_{3}$ and (Li/Na/Mg)[(BN)$_{2}$OBCl$_{3}$] (g-j), respectively.
Up and down spin bands are colored differently. 
Energies are relative to the Fermi energy (red line).
}
\label{Bands}
\end{figure*}

Figs. \ref{Structures} and \ref{StructuresBCl3} depict the structures of 
(BN)$_{2}$OSO$_{2}$F, Li[(BN)$_{2}$OBF$_{3}$] and (BN)$_{2}$OBCl$_{3}$ and its
Li, Na or Mg reduced versions, with one Li, Na or Mg atom per formula unit. 
Images of (BN)$_{2}$OBF$_{3}$ and its corresponding Na and Mg reduced
versions can be found in our former work, Ref.
\cite{nemeth2018simultaneous}. The band structures of h-BN, (BN)$_{2}$OSO$_{2}$F,
(BN)$_{2}$OBF$_{3}$, (BN)$_{2}$OBCl$_{3}$ and the Li, Na and Mg reduced versions of
the latter two ones are compared in Fig. \ref{Bands}. Table \ref{electronicprop} lists
electronic structure properties (band gaps, magnetizations, L{\"o}wdin charges) 
while Tables \ref{bondlengths} and \ref{bondangles} the 
characteristic bond lengths and angles, respectively.

\begin{table}[h]
\caption{
Calculated electronic structure properties of 
h-BN, (BN)$_{2}$OSO$_{2}$F and A[(BN)$_{2}$OBX$_{3}$] (A=Li,Na,Mg; X=F,Cl).
Band gaps (E$_{g}$), total (tot) and absolute (abs) magnetizations (M) 
and L{\"o}wdin charges (Q) are shown. 
The range of Q values is given in e/100 units for B and -e/100 units for N. 
Band gaps of direct transitions are shown. 
(BN)$_{2}$OBF$_{3}$ systems are antiferromagnetic, while (BN)$_{2}$OBCl$_{3}$
and (BN)$_{2}$OSO$_{2}$F ones are ferromagnetic.
}
\label{electronicprop}
\begin{tabular}{lccccc}
\hline
material   &   E$_{g}$      & \multicolumn{2}{c}{M ($\mu_{B}$)} & \multicolumn{2}{c}{Q (e/100) }  \\
           &   (eV)         & tot                      & abs    & B(+)  & N(-)     \\ \hline                                                                                                     
 h-BN                       &   4.7 &  0.0   &  0.0    &  47     &  40       \\
 (BN)$_{2}$OSO$_{2}$F       &  0.0  &  1.75  &  1.97   & 24-50   & 26-40     \\
 (BN)$_{2}$OBF$_{3}$        &  0.8  &  0.13  &  3.28   & 29-53   & 18-25     \\
 Li[(BN)$_{2}$OBF$_{3}$]    &  0.0  &  0.01  &  1.53   & -4-51   & 13-48     \\
 Na[(BN)$_{2}$OBF$_{3}$]    &  0.0  &  0.02  &  1.50   & 3-47    & 13-37     \\
 Mg[(BN)$_{2}$OBF$_{3}$]    &  3.4  &  0.00  &  0.00   & 34-40   & 40-50     \\
 (BN)$_{2}$OBCl$_{3}$       &  0.0  &  2.08  &  2.25   & 14-54   & 24-36     \\
 Li[(BN)$_{2}$OBCl$_{3}$]   &  0.0  &  2.04  &  2.24   & -4-51   & 12-49     \\
 Na[(BN)$_{2}$OBCl$_{3}$]   &  0.0  &  2.02  &  2.22   &  5-47   & 12-34     \\
 Mg[(BN)$_{2}$OBCl$_{3}$]   &  3.9  &  0.00  &  0.00   & 35-41   & 38-50     \\
\hline
\end{tabular}
\end{table}

\begin{table}[h]
\caption{
Calculated characteristic bond lengths in 
h-BN, (BN)$_{2}$OSO$_{2}$F and A[(BN)$_{2}$OBX$_{3}$] (A=Li,Na,Mg; X=F,Cl).
The shorter B-O bond is always the
one within the OBX$_{3}$ unit. 
There is an intramolecular fluorine transfer in (BN)$_{2}$OBF$_{3}$ 
which has a structure of (BNF)(BNOBF$_{2}$) with the shorter B-F bond in the 
OBF$_{2}$ unit.
}
\label{bondlengths}
\begin{tabular}{lccc}
\hline
material   &   \multicolumn{3}{c}{bond lengths ({\AA})}  \\
           &    B-N        & B-O        & B-X  \\
\hline                                                                                                     
 h-BN                       & 1.45       &        --  & --              \\
 (BN)$_{2}$OSO$_{2}$F       & 1.50-1.60  &  --,  1.53 & --              \\
 (BN)$_{2}$OBF$_{3}$        & 1.58-1.61  & 1.37, 1.44 & 1.33, 1.41      \\
 Li[(BN)$_{2}$OBF$_{3}$]    & 1.48-1.59  & 1.44, 1.46 & 1.42-1.45       \\
 Na[(BN)$_{2}$OBF$_{3}$]    & 1.49-1.60  & 1.43, 1.45 & 1.42-1.45       \\
 Mg[(BN)$_{2}$OBF$_{3}$]    & 1.49-1.59  & 1.40, 1.58 & 1.37,1.48       \\
 (BN)$_{2}$OBCl$_{3}$       & 1.47-1.58  & 1.34, 1.49 & 1.86-2.02      \\
 Li[(BN)$_{2}$OBCl$_{3}$]   & 1.50-1.60  & 1.38, 1.48 & 1.85-1.98      \\
 Na[(BN)$_{2}$OBCl$_{3}$]   & 1.49-1.59  & 1.36, 1.50 & 1.82-2.23      \\
 Mg[(BN)$_{2}$OBCl$_{3}$]   & 1.49-1.62  & 1.42, 1.55 & 1.80-1.92      \\
\hline
\end{tabular}
\end{table}

\begin{table}[h]
\caption{
Calculated characteristic bond angles
h-BN, (BN)$_{2}$OSO$_{2}$F and A[(BN)$_{2}$OBX$_{3}$] (A=Li,Na,Mg; X=F,Cl).
}
\label{bondangles}
\begin{tabular}{lcc}
\hline
material   &   \multicolumn{2}{c}{bond angles (deg)}  \\
           &   N-B-O  & B-O-(B,S)   \\
\hline                                                                                                     
 h-BN                          &   --     &   --       \\
 (BN)$_{2}$OSO$_{2}$F          &  99-109  &  131-133   \\
 (BN)$_{2}$OBF$_{3}$           &  99-110  &  143-146   \\
 Li[(BN)$_{2}$OBF$_{3}$]       &  105-111 &   135      \\
 Na[(BN)$_{2}$OBF$_{3}$]       &  101-109 &   136      \\
 Mg[(BN)$_{2}$OBF$_{3}$]       &  100-110 &  138-140   \\
 (BN)$_{2}$OBCl$_{3}$          &  105-111 &  134-149  \\
 Li[(BN)$_{2}$OBCl$_{3}$]      &  101-110 &  135-136  \\
 Na[(BN)$_{2}$OBCl$_{3}$]      &  102-110 &  132-133  \\
 Mg[(BN)$_{2}$OBCl$_{3}$]      &  101-115 &  132-133  \\
\hline
\end{tabular}
\end{table}

Figs. \ref{Bands} a and b indicate the band structures of h-BN and
(BN)$_{2}$OSO$_{2}$F, respectively. The large calculated direct band gap of h-BN of 4.7 eV
(experimental value 5.971 eV \cite{watanabe2004direct}) is reduced to zero upon
functionalization with the $\cdot$OSO$_{2}$F radical in (BN)$_{2}$OSO$_{2}$F.
This is in qualitative agreement with the observed metallic conductivity of 1.5
S/cm in (BN)$_{2}$OSO$_{2}$F \cite{shen1999intercalation}.

Calculations indicate ferromagnetic ground state in (BN)$_{2}$OSO$_{2}$F and in
the (BN)$_{2}$OBCl$_{3}$ systems, while antiferromagnetic ground state has been
predicted for the (BN)$_{2}$OBF$_{3}$ systems. Magnetic susceptibility measurements
on (BN)$_{2}$OSO$_{2}$F found a paramagnetic behaviour \cite{shen1999intercalation}, 
often seen in open shell systems with unpaired electrons. This is in agreement with
the expectation that strongly oxidizing (electron withdrawing) radicals functionalizing 
the B sites create holes (electron deficiency) in the neighboring N atoms, decrease the
electron density in the lone pairs of the N atoms and in the B-N bonds \cite{nemeth2018simultaneous}. 
This effect is also seen on the increased B-N bondlengths and strong deviation 
from planarity in the bond angles as shown in Table \ref{bondangles}.
The atomic charges indicate similar tendency, the B and N atoms lose significant
electron density upon functionalization. Upon reduction with Li, Na and Mg, the electron
densities get partially restored, especially in the case of Mg reduction, as one Mg per
formula unit donates two electrons. 

Note that the valence shell of the B atom is electron
deficient by two electrons in h-BN \cite{nemeth2018simultaneous}, 
therefore each formula unit of (BN)$_{2}$OSO$_{2}$F could in principle take up three
additional electrons, as one electron is provided by the functionalizing $\cdot$OSO$_{2}$F 
radical. For simplicity of notation, X is used in the following, whenever a general
reference to a halide, such as F or Cl, is made in these systems. 
(BN)$_{2}$OBX$_{3}$ could in principle take up four electrons, 
as the B-O bond in the functionalizing $\cdot$OBX$_{3}$ radical can also take up an additional 
electron, as opposed to the S in the $\cdot$OSO$_{2}$F which is saturated with valence +6. 

In practice, only two electrons per formula unit of (BN)$_{2}$OBX$_{3}$ can be taken up, 
each going formally into the two electron deficient B-O bonds in the B-O-B links between   
the h-BN substrate and the $\cdot$OBX$_{3}$ radical. The
(BN)$_{2}$OBX$_{3}$ system can also be viewed as an oxygen double radical
sandwiched between a B center of the h-BN monolayer 
and the B of a neighboring BX$_{3}$ molecule, the latter being a strong
Lewis acid. Once two electrons are taken up by the B-O-B links, 
the non-functionalized B atoms of (BN)$_{2}$OBX$_{3}$ are not
expected to absorb additional electrons in practice, as they are not strong enough Lewis acids, 
even though their valence shells are still incomplete by two electrons.
More realistically, the electron deficient B-O-BX$_{3}$ links withdraw electrons from the
nearby N atoms creating holes in them as soon as the functionalization is established 
and these N-holes will take up the electrons later provided by reductive Li, Na or Mg atoms or
by any other suitable reductive agents. After the reduction of (BN)$_{2}$OBX$_{3}$, the
extra one or two electrons will largely be localized on the B-O bonds in the B-O-BX$_{3}$
links making these links once or twice negatively charged ions.

Some B-N bond-lengths in the (BN)$_{2}$OBX$_{3}$ systems become as long as 1.62 {\AA},
which is a substantial deviation from 1.45 {\AA} seen in h-BN. However, such long B-N bonds
are not unusual as the B-N bondlength in cubic BN is 1.57 {\AA} (sphalerite type) or
1.55-1.58 {\AA} (wurtzite type), 1.58 {\AA} in ammonia borane (H$_{3}$N$\rightarrow$BH$_{3}$) and
1.66 {\AA} in H$_{3}$N$\rightarrow$BF$_{3}$. 

One possible application of the (BN)$_{2}$OBX$_{3}$ systems is as cathode active species
in electrochemical energy storage devices, such as batteries and super/pseudo-capacitors. In
such applications, the reduction of the (BN)$_{2}$OBX$_{3}$ species happens during
discharge, for example as
\begin{equation}
\rm
\label{discharge1}
Li^{+} + e^{-} + (BN)_{2}OBX_{3} \rightleftharpoons Li[(BN)_{2}OBX_{3}] , \\
\end{equation}
and
\begin{equation}
\rm
\label{discharge2}
Li^{+} + e^{-} + Li[(BN)_{2}OBX_{3}] \rightleftharpoons Li_{2}[(BN)_{2}OBX_{3}] .   
\end{equation}
In the present work, only one Li, Na or Mg atom is used to reduce
the (BN)$_{2}$OBX$_{3}$ materials per formula unit, however, in principle up to two Li or
Na atoms or one Mg atom per formula unit can be absorbed to provide the two electrons that the
above mentioned B-O-B links and the related N-holes can absorb. The reason for investigating
only one reductive atom per formula unit here is that the first reductive atom can be
placed within the monolayer between the functional groups, while the 
second atom would be placed between the monolayers and would require additional complexity of
modeling that we will consider in a separate study. Nonetheless, it is worth mentioning that
the absorption of a second electron per formula unit is expected to 
near double the capacity and energy
density of the Li and Na based discharge processes. For Mg, the two electron absorption
process is discussed here as the Mg$^{2+}$ ion fits into the monolayer.
The calculated open circuit voltages, gravimetric capacities and energy densities 
of the corresponding cell reactions are listed in 
Table \ref{electrochem}. 

\begin{table}[h]
\caption{Calculated electrochemical properties of 
(BN)$_{2}$OBX$_{3}$ (X=F,Cl) with Li, Na
and Mg anodes. Open circuit voltage (OCV), 
gravimetric capacity (GC) and energy density (GED) values are shown.
Discharged states investigated here contain one anode atom per formula unit, which is half 
of the    
theoretical maximum capacity of (BN)$_{2}$OBX$_{3}$ 
for the Li and Na systems and full capacity for Mg.
}
\label{electrochem}
\begin{tabular}{lccc}
\hline
                         & OCV  &  GC    &   GED    \\
discharged state         & (V)  &(mAh/g) & (Wh/kg)  \\
\hline                                           
Li[(BN)$_{2}$OBF$_{3}$]  & 5.60 & 191    & 1067     \\
Na[(BN)$_{2}$OBF$_{3}$]  & 5.10 & 171    &  874     \\
Mg[(BN)$_{2}$OBF$_{3}$]  & 3.60 & 340    & 1222     \\
Li[(BN)$_{2}$OBCl$_{3}$] & 3.69 & 141    &  521     \\
Na[(BN)$_{2}$OBCl$_{3}$] & 3.44 & 130    &  448     \\
Mg[(BN)$_{2}$OBCl$_{3}$] & 2.78 & 259    &  719     \\
\hline
\end{tabular}
\end{table}

The most energetic cell reaction happens between a Li anode and $\cdot$OBF$_{3}$ functionalized
h-BN at a voltage of 5.6 V with a capacity of 191 mAh/g and energy density of 1067 Wh/kg. The
voltage for this cathode material decreases to 5.1 V for Na anode and so does the capacity
(171 mAh/g) and the energy density (874 Wh/kg) as well. With Mg anode, two
electrons are absorbed, thus albeit the voltage decreases to 3.6 V, the capacity becomes 340 mAh/g
and the energy density 1222 Wh/kg.  Exchanging the F in the cathode to Cl still produces very
energetic cell reactions with voltages of 3.69, 3.44 and 2.78 V while capacities are 141, 130
and 259 mAh/g and energy densities 521, 448 and 719 Wh/kg, for Li, Na and Mg anodes (only one anode
atom absorbed per formula unit), respectively.

The band gaps of the reduced systems are zero, as long as there is a sufficient concentration
of N-holes in the system, i.e. until significantly less than two electrons per formula unit
are absorbed. In the calculated discharged systems, the Li and Na discharged species still have
zero band gap, while the Mg ones have a 3.4-3.9 eV band gap.
The size of the band gap is informative about the expected
electronic conductivity, albeit the latter also depends on the mobility of the charge carriers.
Since the charge carriers are the N-holes in these systems, the electronic conductivity is
likely to be similar to the baseline material (BN)$_{2}$OSO$_{2}$F
and be in the order of 1.5 S/cm at room temperature 
\cite{shen1999intercalation}. As mentioned above, this
electronic conductivity is in the same order with practical graphene 
(having lattice defects) and with carbon black conductive additives used in batteries.

The proton conductivity of functionalized h-BN is in the order of 0.1 S/cm at room
temperature \cite{mofakhamipatents} and similarly great Li ion conductivities can be expected
in the A$_{n}$[(BN)$_{2}$OBX$_{3}$] materials proposed here, 
where A is an alkali atom (Li, Na, etc; 0$\le$n$\le$2) or alkaline earth (Mg, etc; 0$\le$n$\le$1). 
This property will be discussed below in a separate section.

The combined features of the A$_{n}$[(BN)$_{2}$OBX$_{3}$] compounds allow for many
valuable functions by a single material, such as electroactive species, solid ion conductors
and potentially even as electroconductive material, at the same time. 
Reduction of the number of different phases in a (solid state) battery can be very helpful,
for example for reducing interfacial impedance \cite{pervez2019interface}.
It is expected that 
A$_{n}$[(BN)$_{2}$OBX$_{3}$] are also stable at elevated temperatures and safe as only
solid decomposition products are expected from them upon high temperature heating. These
features will be discussed in forthcoming sections.

\section{Stability of functionalized \lowercase{h}-BN-s}
\subsection{Stability of $\cdot$OSO$_{2}$F functionalized h-BN}
As described in Ref \onlinecite{shen1999intercalation}, $\cdot$OSO$_{2}$F
functionalized h-BN was stable over several months, unless it came in contact
with base metals, in which case it quickly decomposed. Decomposition also
happened when liquid HF, dried over K$_{2}$NiF$_{6}$ was distilled 
over samples of $\cdot$OSO$_{2}$F 
functionalized h-BN. No decomposition happened with
HF gas, indicating that traces of K$_{2}$NiF$_{6}$, a transition metal salt,
might be responsible for the decomposition of the h-BN derivative. 
Also, quantum chemical calculations in the present work indicate the
detachment of [OSO$_{2}$F]$^{-}$ ions from h-BN, when (BN)$_{2}$OSO$_{2}$F
is lithiated.
The decomposition of a sample with (BN)$_{3}$OSO$_{2}$F stoichiometry 
was precisely measured \cite{shen1999intercalation} to happen according to 
\begin{equation}
\label{Decomp1}
\begin{aligned}
\rm 6  (BN)_{3}OSO_{2}F \rightleftharpoons & \rm 17 BN + 5/2 S_{2}O_{5}F_{2} + & \\   
\rm                                    & \rm + NO[OSO_{2}F] + 1/2 B_{2}O_{3} , &
\end{aligned}
\end{equation}
where S$_{2}$O$_{5}$F$_{2}$ has the structure of FSO$_{2}$-O-SO$_{2}$F with a single
oxygen atom between the fluorosulfonyl groups 
and NO[OSO$_{2}$F] consists of a NO$^{+}$ (nitrosonium) cation
and a non-radical [OSO$_{2}$F]$^{-}$ anion formed from the charge neutral radical
$\cdot$OSO$_{2}$F after the absorption of an electron. The precise mechanism of this 
decomposition is unknown. The present author speculates, that it is catalytic
amounts of dissolved transition metals that induce the detachment of
$\cdot$OSO$_{2}$F radicals from h-BN in the reduced form of [OSO$_{2}$F]$^{-}$
anions by transferring six electrons to these radicals, and on turn the transition
metals recover these six electrons from the oxidation of a nitride ion of h-BN from
oxidation number -3 to +3 resulting in the formation of a nitrosonium ion. 
Since the process of Eq \ref{Decomp1} also involves the detachment of oxide ions
from the [OSO$_{2}$F]$^{-}$ ions before S$_{2}$O$_{5}$F$_{2}$ forms, the process may
also include the temporary valence change of the sulfur atoms.
This speculated process may formally be partitioned into the following 
equations without claiming that they would describe the precise mechanism:
\begin{eqnarray}
\rm
\label{Decomp2}
6 (BN)_{3}OSO_{2}F + 6 e^{-} \rightleftharpoons 18 BN + 6 [OSO_{2}F]^{-},\\
\rm
\label{Decomp3}
5 [OSO_{2}F]^{-} \rightleftharpoons  5/2 O^{2-} + 5/2 S_{2}O_{5}F_{2}, \\
\rm
\label{Decomp4}
BN + 5/2 O^{2-} \rightleftharpoons  1/2 B_{2}O_{3} + NO^{+} + 6 e^{-}.   
\end{eqnarray}
Since one of the reaction products of Eq \ref{Decomp1}, namely
S$_{2}$O$_{5}$F$_{2}$, is volatile and is removed from the system, the equilibrium
is shifted toward the complete decomposition of (BN)$_{3}$OSO$_{2}$F. Without a
volatile product, the decomposition is expected to be slow if happening at all.

\subsection{Stability of $\cdot$OBF$_{3}^{-}$ functionalized h-BN}
\subsubsection{In the discharged state}
Assuming that the decomposition of Li[(BN)$_{2}$OBF$_{3}$] would happen similarly to
that of Eq \ref{Decomp1}, the analogous reaction would follow as:
\begin{equation}
\label{Decomp5}
\begin{aligned}
\rm 
6 Li[(BN)_{2}OBF_{3}] & \rm \rightleftharpoons 11 BN + 5/2 Li_{2}[F_{3}B-O-BF_{3}] + & \\
\rm
                      & \rm + LiNO[OBF_{3}] + 1/2 B_{2}O_{3} , &
\end{aligned}
\end{equation}
where [F$_{3}$B-O-BF$_{3}$]$^{2-}$ is a Lewis adduct of an oxide ion 
with two BF$_{3}$ molecules and is an anionic ether, 
similar to the charge neutral ether FSO$_{2}$-O-SO$_{2}$F in Eq \ref{Decomp1}. 
LiNO[OBF$_{3}$] is a salt of the [OBF$_{3}$]$^{2-}$
anion with one NO$^{+}$ and one Li$^{+}$ cation. Both anions are non-radical ones
and are also analogous to the well known BF$_{4}^{-}$ anion with one F$^{-}$ exchanged to
an O$^{2-}$. Also note that closely related nitrosonium tetrafluoroborate, 
NO[BF$_{4}$] is a well known compound.
All products of reaction Eq \ref{Decomp5} 
$-$ beside h-BN $-$ are salts and are non-volatile. Therefore,
it is expected that the decomposition of Li[(BN)$_{2}$OBF$_{3}$] would be much slower
than that of (BN)$_{3}$OSO$_{2}$F. Also the lack of a second valence changing atom $-$ 
besides nitrogen $-$ may hinder the decomposition: the structural equivalent of sulfur of
$\cdot$OSO$_{2}$F would be boron of $\cdot$OBF$_{3}^{-}$, which is not expected
to change its valence in the given circumstances. Electrochemical extraction
of the Li$^{+}$ ions (charging of the battery) may potentially even reverse this
decomposition and produce the charged state of the system.  
At elevated temperatures the following additional decompositions may happen:
\begin{eqnarray}
\label{Decomp6}
\rm LiNO[OBF_{3}] & \rm \rightleftharpoons LiNO_{2} + BF_{3}, \\ 
\label{Decomp7}
\rm Li_{2}[F_{3}B-O-BF_{3}] & \rm \rightleftharpoons Li_{2}O + 2 BF_{3},  
\end{eqnarray}
producing the only gaseous decomposition product, BF$_{3}$. 
However, for such a decomposition to happen, ionic Lewis adducts must break up 
which is expected to require high temperatures similar to the 
decomposition temperature of LiBF$_{4}$ to LiF and BF$_{3}$ which happens
between 162 and 277 $^{o}$C with peak at 263 $^{o}$C \cite{lu2006thermal}.
Even further, BF$_{3}$ can react with B$_{2}$O$_{3}$ to form 
solid boron-oxyfluoride, BOF:
\begin{equation}
\rm
BF_{3} + B_{2}O_{3} \rightleftharpoons 3 BOF .
\end{equation}
BOF may have several different structures, such as glass or six-membered
rings and is stable even at very high temperatures 
\cite{polishchuk2011oxyfluoride,boussard1997structure}.

\subsubsection{In the charged state}
The fully charged state has the chemical formula (BN)$_{2}$OBF$_{3}$.
In this state, the single radical $\cdot$OBF$_{3}^{-}$ anion
is oxidized to a charge neutral double radical {:}OBF$_{3}$ 
which is the Lewis adduct of a double radical oxygen atom with BF$_{3}$. Then the charge
neutral double radical {:}OBF$_{3}$ splits into two single radicals,
according to theoretical predictions: one $\cdot$F and one $\cdot$OBF$_{2}$
\cite{nemeth2018simultaneous}. Each of these radicals functionalize separate
nearby boron atoms as expressed by the formula (BNF)(BNOBF$_{2}$). Note that such a splitting
of the {:}OBCl$_{3}$ radical has not been seen in the present calculations.
Assuming a similar decomposition as in the above cases, the following is
expected:
\begin{equation}
\label{Decomp8}
\begin{aligned}
\rm 
6 [(BN)_{2}OBF_{3})] & \rm \rightleftharpoons 11 BN + 5/2 F_{3}B-O-BF_{3} + & \\
\rm
                      & \rm + NO[OBF_{3}] + 1/2 B_{2}O_{3} , &
\end{aligned}
\end{equation}
where F$_{3}$B-O-BF$_{3}$ is the Lewis adduct of an oxygen atom with two
BF$_{3}$ molecules and NO[OBF$_{3}$] is the salt of a nitrosonium cation
with a $\cdot$OBF$_{3}^{-}$ radical anion. F$_{3}$B-O-BF$_{3}$ may further
decompose according to
\begin{equation}
\label{Decomp9}
\rm 
F_{3}B-O-BF_{3} \rightleftharpoons BF_{3} + FOBF_{2} .
\end{equation}
Further, BF$_{3}$ gas may be bound through reaction with B$_{2}$O$_{3}$ 
to form solid BOF, as mentioned also above.
Therefore, it is expected that most decomposition products will be solids
even at elevated temperatures.

\section{Synthetic strategies}
In the following subsections, synthetic strategies for the production of (BN)$_{2}$OBF$_{3}$
and its reduced states will be discussed. Analogous strategies can be applied for 
(BN)$_{2}$OBCl$_{3}$ and its reduced versions as well.

\subsection{Charged state: (BN)$_{2}$OBF$_{3}$}
Perhaps the most efficient synthesis for the charged state of the
proposed (BN)$_{2}$OBF$_{3}$ electroactive species is based on reacting h-BN
with a mixture of ozone (O$_{3}$) and BF$_{3}$ in an inert 
medium, as proposed in our former work Ref \onlinecite{nemeth2017OzoneBF3}. 
An ideal such medium for ozonation reactions is liquid CO$_{2}$ in which high
concentrations of ozone can be used safely 
\cite{ozonolysisBala2013,lundin2015liquid}.

In a similar, but somewhat simpler setting proposed here, 
a mixture of concentrated
H$_{2}$O$_{2}$ and a source of BF$_{3}$ \cite{mcclure1962hydrogen}, 
such as ether solution of BF$_{3}$  
or BF$_{3}$ gas may be used to generate Lewis adducts 
H$_{2}$O$_{2}$.2BF$_{3}$ which spontaneously split to $\cdot$OHBF$_{3}$
radicals and these radicals covalently bind to the boron atoms of h-BN. The
protons can be removed electrochemically to obtain (BN)$_{2}$OBF$_{3}$.
Another equivalent of this latter process is to functionalize h-BN with
$\cdot$OH radicals first, then react the product with BF$_{3}$ 
\cite{nemeth2019radicalAnion} and extract
the protons electrochemically.

\subsection{Radical anion functionalization 
to obtain the discharged state: Li[(BN)$_{2}$OBF$_{3}$]} \label{RadicalAnionFunct}

The robust and fast synthesis of (BN)$_{2}$OSO$_{2}$F carried out first in 1978
\cite{bartlett1978novel,shen1999intercalation}, using liquid peroxide
FSO$_{2}$-O-O-SO$_{2}$F as a source of $\cdot$OSO$_{2}$F radicals, 
has inspired the present author to look for a similar peroxide based synthesis
opportunity for the discharged state, Li[(BN)$_{2}$OBF$_{3}$], 
of the proposed cathode active material. 
By deleting the BN content of the formula Li[(BN)$_{2}$OBF$_{3}$], the remaining        
species would form Li[OBF$_{3}$] containing a Li$^{+}$
cation and a $\cdot$OBF$_{3}^{-}$ radical anion. Such a radical anion can formally
be derived from a salt Li$_{2}$[F$_{3}$B-O-O-BF$_{3}$] by splitting the
peroxide bond in its [F$_{3}$B-O-O-BF$_{3}$]$^{2-}$ anion. Since the
BF$_{3}$ unit comes ultimately from BF$_{3}$ gas, this salt may be composed as the
Lewis adduct of lithium peroxide Li$_{2}$O$_{2}$ with BF$_{3}$:
\begin{equation}
\label{LewisAdduct1}
\rm
Li_{2}O_{2} + 2 BF_{3} \rightleftharpoons Li_{2}[F_{3}B-O-O-BF_{3}] .
\end{equation}
It is expected that melting Li$_{2}$[F$_{3}$B-O-O-BF$_{3}$] would
lead to -O-O- bond splitting and the resulting radical anions would
functionalize h-BN similarly to liquid FSO$_{2}$-O-O-SO$_{2}$F
and the $\cdot$OSO$_{2}$F radicals \cite{nemeth2019radicalAnion}:
\begin{eqnarray}
\rm
\label{Decomp0}
[F_{3}B-O-O-BF_{3}]^{2-} \rightleftharpoons 2 [{\cdot}OBF_{3}]^{-} \\
\rm
2 Li^{+} + 2 [{\cdot}OBF_{3}]^{-} + 4 BN \rightleftharpoons 2 Li[(BN)_{2}OBF_{3}] .
\end{eqnarray}
Lewis adducts, including adducts with BF$_{3}$, 
have been studied in past years as additives to liquid electrolytes
that improve the performance of Li ion batteries 
\cite{nie2016some,xiao2016lewisadducts,matsui2014,wang2000}.
In principle, the surface of cathode active species, such as the surface of LiCoO$_{2}$,
can irreversibly bind BF$_{3}$ and other Lewis acids, as such binding has been
known for Al$_{2}$O$_{3}$, SiO$_{2}$, TiO$_{2}$, ZrO$_{2}$, NiO, etc, 
resulting in M-OBF$_{2}$, M-OBF$_{3}^{-}$ and (M)-O-(BF)-O-(M) 
surface functionalizations (M metal)
\cite{BF3Al2O3Patent1947,rhee1968chemisorption,morrow1976infrared,sadeghi2008bf}. 
The binding of Lewis acids to Lewis base surface sites may potentially be 
utilized in the construction of artificial
solid-electrolyte interfaces.

Since the fluoride (halide) ions bound to boron have a great mobility
in some solvents (such as in liquid boron trihalides) \cite{hartman1978adducts}
the reaction of Li$_{2}$O$_{2}$ with BF$_{3}$ may have some alternative
products as well, depending on the solvent and reaction circumstances. An
alternative product would be a mixture of cyclic perborate
Li$_{2}$[F$_{2}$B(-O-O-)$_{2}$BF$_{2}$] and LiBF$_{4}$: 
\begin{equation}
\label{LewisAdduct2}
\rm
2 Li_{2}O_{2} + 4 BF_{3} \rightleftharpoons 
Li_{2}[F_{2}B(-O-O-)_{2}BF_{2}] + 2 LiBF_{4} .
\end{equation}
Such cyclic perborates are well known, for example sodium perborate 
Na$_{2}$[(OH)$_{2}$B(-O-O-)$_{2}$B(OH)$_{2}$] is produced on the industrial scale
for use as a bleaching agent. The thermal decomposition of sodium perborate
has been studied and 
interpreted to go through the simultaneous splitting of both peroxide bonds to form
radical anion species [O$_{2}$B(OH)$_{2}]^{-}$ that is followed by the release of
water and formation of sodium metaborate \cite{koga2018multistep}:
\begin{eqnarray}
\label{LewisAdduct3}
\rm
[(OH)_{2}B(-O-O-)_{2}B(OH)_{2}]^{2-} \rightleftharpoons  
2 [({\cdot}O)_{2}B(OH)_{2}]^{-} \\
\label{LewisAdduct4}
\rm
2 Na^{+} + 2 [({\cdot}O)_{2}B(OH)_{2}]^{-} \rightleftharpoons 
2 NaBO_{2} + O_{2} + 2 H_{2}O .
\end{eqnarray}
Obviously, the analogous reaction with F instead of OH groups would not happen as it
should release F$_{2}$ molecules instead of O$_{2}$ and H$_{2}$O, but F$_{2}$ is
extremely reactive. Instead, the formation of BOF could be expected:
\begin{eqnarray}
\label{LewisAdduct6}
\rm
[F_{2}B(-O-O-)_{2}BF_{2}]^{2-} \rightleftharpoons  
2 [({\cdot}O)_{2}BF_{2}]^{-} \\
\label{LewisAdduct7}
\rm
2 Li^{+} + 2 [({\cdot}O)_{2}BF_{2}]^{-} \rightleftharpoons 
2 LiF + O_{2} + 2BOF  .
\end{eqnarray}
In light of the latter decomposition process, one must be careful to raise the
temperature only up to the point of breaking the peroxide bonds (Eqns \ref{Decomp0} 
and \ref{LewisAdduct6}), but not to release
O$_{2}$, so that the [($\cdot$O)$_{2}$BF$_{2}$]$^{-}$ and $\cdot$OBF$_{3}^{-}$
radical anions can functionalize h-BN.
This latter condition is however not surprising, it has been the same for the 
old, $\cdot$OSO$_{2}$F functionalization of h-BN: would the temperature be too high,
an explosive decomposition of the reagent FSO$_{2}$-O-O-SO$_{2}$F could happen:
\begin{equation}
\rm
2 FSO_{2}-O-O-SO_{2}F \rightleftharpoons 2 FSO_{2}-O-SO_{2}F + O_{2} .
\end{equation}
However, when carefully handled, as in Refs \onlinecite{bartlett1978novel} and 
\onlinecite{shen1999intercalation}, even the distillation of FSO$_{2}$-O-O-SO$_{2}$F
could be carried out without explosive decomposition. Therefore, it is expected that
the Lewis adducts of Li$_{2}$O$_{2}$ with BF$_{3}$ will also be stable in a
temperature range above their melting points and can generate stable radical anions
for the functionalization of h-BN. Both above mentioned radical anions
$\cdot$OBF$_{3}^{-}$ and
[($\cdot$O)$_{2}$BF$_{2}$]$^{-}$ are suitable to functionalize h-BN and both products 
can be used as cathode active species in batteries. Also note that 
[($\cdot$O)$_{2}$BF$_{2}$]$^{-}$ radical anions will occur in pair with BF$_{4}^{-}$
as pointed out in Eq \ref{LewisAdduct2}. This does not hinder the
functionalization of h-BN,
or the use of the product of the functionalization as 
cathode active species: BF$_{4}^{-}$ ions would be part of the
(solid) electrolyte surrounding the functionalized h-BN.

\subsection{Radical anions from H$_{2}$O.BF$_{3}$ derivatives}
An alternative source of $\cdot$OBF$_{3}^{-}$ radical anions may be based on the
well known Lewis adduct of water with BF$_{3}$, H$_{2}$O.BF$_{3}$, and its
salts, such as LiOH.BF$_{3}$ and Li$_{2}$O.BF$_{3}$. Electrochemical extraction of 
one hydrogen or one alkali atom per formula unit results in 
salts (LiOBF$_{3}$) and Br{\o}nsted acids (HOBF$_{3}$) containing the desired
radical anion. Therefore, a mixing of these derivatives of
H$_{2}$O.BF$_{3}$ with h-BN and electrolysis of the mixture will likely
result in $\cdot$OBF$_{3}^{-}$-functionalized h-BN. While NaHOBF$_{3}$
has been synthesized by the reaction of NaHCO$_{3}$ and BF$_{3}$ in aqueous
solution \cite{Gilpatrick1974,clark1970crystal}, the analogous LiOBF$_{3}$
has not been synthesized yet \cite{Zhang2015KineticSO}. Nothing is known of 
Li$_{2}$OBF$_{3}$ either. H$_{2}$O.BF$_{3}$ itself hydrolyzes in
water to boric acid, HF and a number of other intermediates 
\cite{wamser1951equilibria}. Therefore, it appears that routes to
$\cdot$OBF$_{3}^{-}$ based on derivatives of H$_{2}$O.BF$_{3}$ are less
certain and less efficient than those based on peroxide bond splitting, 
except perhaps the route based on NaHOBF$_{3}$ which can be synthesized
in a very economic fashion.

\subsection{Radical anions other than $\cdot$OBF$_{3}^{-}$}
The functionalization of 2D materials with general radical anions
other than $\cdot$OBF$_{3}^{-}$ is entirely possible. One such
possibility is through splitting the N-N bond in Lewis adducts of
hydrazine and BF$_{3}$. Such adducts with H$_{4}$N$_{2}$.BF$_{3}$ or 
H$_{4}$N$_{2}$.2BF$_{3}$ composition are known, they form acidic solutions
in water even before hydrolysis \cite{paterson1961interaction}.
The splitting of the N-N bond of these compounds results in 
radical anions, such as [$\cdot$NHBF$_{3}$]$^{-}$ and protons.
Functionalization of h-BN with these radical anions produces
Br{\o}nsted acid functionalized h-BN. The N-protons can 
at least partially be exchanged to Li$^{+}$ or other cations  
electrochemically and be used as the discharged state of electroactive
species. 

\section{Optimizing the extent of the discharge}
In principle, absorption of up to two electrons per formula unit of (BN)$_{2}$OBX$_{3}$
is possible, therefore the general formula for the discharged state with Li anode and
(BN)$_{2}$OBF$_{3}$ cathode is Li$_{n}$[(BN)$_{2}$OBF$_{3}$] with 0$<$n$\le$2.
For discharge 1$<$n$\le$2, however, the radical character of the functionalizing 
$\cdot$OBF$_{3}^{-}$ species will be lost when they pick up an additional electron
and become [OBF$_{3}$]$^{2-}$. Once the radical character is lost, the
anion may still be bound to the boron atom of h-BN if there is a
sufficiently strong Lewis base (anion) and Lewis acid (B of h-BN) 
interaction between them. Unfortunately, 
the boron atom of a planar h-BN is a too week Lewis acid as the planarity is
strongly preferred due to the short B-N bonds. However, the detachment of
[OBF$_{3}$]$^{2-}$ anions is likely not irreversible, as charging of the battery 
re-creates the radical anions. 
Detachment of [OBF$_{3}$]$^{2-}$
is not expected from the edges of the h-BN layers as the B atoms there
have much stronger Lewis acid character than in the middle of the basal
plane. 
The electronic conductivity of the
discharged state for 1$<$n$\le$2 will be much lower though than for 0$<$n$\le$1, 
because the concentration of N-holes decreases as 1$<$n and n$\rightarrow$2
and this leads to greater electronic resistance and less
energy efficient recharge. 
A similar dependence of electronic conductivity on the concentration 
of hole states has been the conclusion of Ref \onlinecite{shen1999intercalation}
on the -OSO$_{2}$F functionalized h-BN.
Clearly, there is a trade-off between increased
energy density and energy efficiency of charging. 

\section{A$_{n}$[(BN)$_{2}$OBX$_{3}$] as a solid electrolyte}
As high as 0.2 mS/cm room temperature Li-ion conductivity has been demonstrated in
h-BN composite pastes containing h-BN, an absorbed ionic liquid electrolyte and a Li-salt
\cite{rodrigues2016hexagonal}. The thermal stability of Li-ion batteries with this
paste electrolyte has also been demonstrated up to 120 $^{o}$C, which is well above
60 $^{o}$C, the upper thermal limit of liquid electrolyte based batteries
\cite{rodrigues2016hexagonal}. While Ref. \onlinecite{rodrigues2016hexagonal}
mentions the interaction of the electrolyte with h-BN, there was no indication of
covalent functionalization of h-BN. Ion gel electrolytes containing 
amine functionalized h-BN show even higher Li-ion conductivity of 0.647 mS/cm at room
temperature \cite{KDonggun2020adfm}.


Similarly great proton-conductivity of h-BN-s functionalized with 
$\cdot$OH, $\cdot$SO$_{4}$H, $\cdot$SO$_{3}$H and $\cdot$NH$_{2}$ radicals
has been experimentally measured and found to be in the order of 0.1 S/cm
when the functionalization was sufficiently dense 
\cite{mofakhami2010material,mofakhami2011material,mofakhami2015material}.
This is a result of
contiguous networks of hydrogen bridges between neighboring functional
groups that allow for quick proton hopping. 
It suggests that
some covalently functionalized h-BN-s may also be well suited for the conduction of other
cations, such as Li$^{+}$, Na$^{+}$, Mg$^{2+}$ and MgX$^{+}$. Radical anion
functionalization appears essential to fulfill this promise as it
establishes a stable covalent binding of negatively charged functional
groups to the surface of h-BN. 

The $\cdot$OBX$_{3}$ functionalized 2D materials can particularly be advantageous 
for the transport of Mg$^{2+}$ ions. In fact, Mg$^{2+}$ ions often travel together with
an attached halide ion \cite{Yao-MgIntercalation-2016}. This attached halide ion may come from
the $\cdot$OBX$_{3}$ units, as the halide ions can quickly and reversibly separate from the boron
centers, especially when the B center is negatively charged and three or four halides are
attached to it \cite{hartman1978adducts}. 
An example of this process with Cl$^{-}$ ions
can be described by the following equation:
\begin{equation}
\rm
\label{Mgtransport}
Mg^{2+} + [(BN)_{2}OBCl_{3}]^{2-} \rightleftharpoons MgCl^{+} + [(BN)_{2}OBCl_{2}]^{-} .
\end{equation}
Therefore, the hopping of MgX$^{+}$ cations between -OBX$_{2}$ sites is expected to
be a fast way of transporting Mg$^{2+}$ ions in the $\cdot$OBX$_{3}$ functionalized h-BN.

The recently discovered paddle wheel mechanism \cite{zhang2020targeting} also supports the use
of the $\cdot$OBX$_{3}$ functional groups for achieving fast ion conduction in solids, as
the rotation of these groups around the O-B bonds promotes the fast motion of cations, similarly to
the Grotthus mechanism of proton transfer that involves a hopping of protons between sites 
and rotations of the protonated sites.

The radical anions of the h-BN surface also
provide a solvation shell for the cations. Even with dense
functionalization,
when every second B atom is functionalized with $\cdot$OBF$_{3}^{-}$, there is
sufficient space between these anions for the quick
hopping of cations between sites. Since the functionalized h-BN 
is intended to be used not only as an ionic conductor, but also as an
electroactive species, the transport of positive ions must be coupled
with the transport of equal number of electrons.
N-holes generated by the electron withdrawing effect of the functionalizing
anionic radicals are the enablers of efficient electronic transport
\cite{nemeth2018simultaneous}. A similar role of O-holes have been
observed in LiCoO$_{2}$, albeit the transport of electrons is
only partially based on O-holes and to a greater extent to valence changes
of cobalt \cite{yoon2002oxygen,mizokawa2013role,seo2016structural,luo2016charge}.

\section{A$_{n}$[(BN)$_{2}$OBX$_{3}$] as an anode coating
and separator}
Graphite fluoride (CF$_{x}$) was recently proposed as a 
coating/packaging material
for lithium metal anodes \cite{shen2019lithium}. It provides well-wetting
surfaces to (solid) electrolytes, sufficient ionic conductivity, inertness to air and  
moisture, and dendrite-free cycling over hundreds of cycles. 
It appears to provide some temperature resistance as well,
considering that the synthesis of the coating was performed at 250 $^{o}$C. Contrary to the
expectation that all the fluorene of CF$_{x}$ should convert to LiF when in
contact with lithium, this conversion does not seem to complete, while some
LiF layers form between Li metal and CF$_{x}$.

Other proposed alternatives to CF$_{x}$ coating include g-C$_{3}$N$_{4}$, which forms a Li$_{3}$N
and graphene layer over the Li metal anode \cite{huang2020graphitic} or a
biomineralization based coating using -O-CH$_{2}$CF$_{3}$ functionalized egg-shell-membrane 
\cite{ju2020biomacromolecules} that allows for stable cycling of a Li metal anode for
3000 cycles at 5 mA/cm$^{2}$ current density.

Also note that highly fluorinated electrolytes have been known for long to allow for
dendrite-free plating of lithium. These electrolytes include fluorosulfonyl imides 
\cite{qian2015high,shkrob2014fluorosulfonyl} and Lewis adducts of BF$_{3}$ with
organic imides and carboxylates \cite{US7534527B2,JP2013209355A,JP2017178964A}.

A$_{n}$[(BN)$_{2}$OBX$_{3}$] is expected to provide a similar coating
on the respective metal anodes, 
except that it would not repel water, however it could potentially provide a greater
thermal stability and ionic conductivity. Also note that BF$_{3}$ is a fire
extinguishing agent for magnesium and, if released by thermal decomposition of
the coating, could potentially help to suppress anode fires.
The main advantage of A$_{n}$[(BN)$_{2}$OBX$_{3}$] coating as compared to
CF$_{x}$ is likely a greatly increased ionic conductivity
since the Li, Na and Mg fluorides have poor ionic conductivities, for example for the small
size of their anions \cite{zhou2020protective}. Much larger anions in
A$_{n}$[(BN)$_{2}$OBX$_{3}$] and a lot of space between the functional groups on the h-BN
monolayers is expected to promote an improved ionic conductivity. 

Similarly to all-solid-state graphene oxide batteries, where graphene oxide plays both the
role of the cathode active species and the separator \cite{ye2017metal}, 
A$_{n}$[(BN)$_{2}$OBX$_{3}$] is also 
expected to be able to function as the separator as well. In this case, the fully reduced and
therefore electronically isolating but ionically conducting 
A$_{n}$[(BN)$_{2}$OBX$_{3}$] material would form the separator between 
the metal anode and the cathode, where n=2 for Li and Na and n=1 for Mg in the fully
reduced states. The fully reduced species would naturally form upon direct
contact of A$_{n}$[(BN)$_{2}$OBX$_{3}$] with the anode. 
If mechanical enforcement is needed,
A$_{n}$[(BN)$_{2}$OBX$_{3}$] can be used as a component of a
polymer-composite for the separator and also in the cathode.

\section{Cost analysis}
Table \ref{costanalysis} lists the materials-only cost of the battery components and the 
cost of storing 1 kWh energy. The theoretical energy densities were taken from
Table \ref{electrochem}. 
It is assumed that A$_{n}$[(BN)$_{2}$OBX$_{3}$] is made of h-BN, BF$_{3}$, BCl$_{3}$ and Li,
Na or Mg metal (for A) and these materials are 
available at the following actual prices of 35, 20, 10, 85, 3.0 and 2.3 USD/kg 
\cite{prices}, respectively (cost of oxygen has been neglected).
It appears, the proposed battery materials are sufficiently cost effective, especially
compared to the current cost of batteries, 156 USD/kWh \cite{trahey2020energy}, albeit labor,
equipment and related costs are not included here.

\begin{table}[h]
\caption{Estimated price of materials and stored energy.}
\label{costanalysis}
\begin{tabular}{lccccc}
\hline
                         &  components   & energy   \\
material                 &    USD/kg     & USD/kWh  \\
\hline                     
Li[(BN)$_{2}$OBF$_{3}$]  &   26          & 24       \\
Na[(BN)$_{2}$OBF$_{3}$]  &   20          & 23       \\
Mg[(BN)$_{2}$OBF$_{3}$]  &   20          & 16       \\
Li[(BN)$_{2}$OBCl$_{3}$] &   18          & 35       \\
Na[(BN)$_{2}$OBCl$_{3}$] &   14          & 32       \\
Mg[(BN)$_{2}$OBCl$_{3}$] &   14          & 20       \\
\hline
\end{tabular}
\end{table}

\section{Summary and Conclusions}
Motivated by a forty years old functionalized h-BN material, (BN)$_{2}$OSO$_{2}$F 
\cite{bartlett1978novel,shen1999intercalation} and its 
good electronic conductivity and instability toward reducing agents as well as by the outstanding
energy and power densities of graphene oxide cathodes 
\cite{jang2011graphene,liu2014lithium,kim2014novel} and the instability of graphene oxide
at elevated temperatures \cite{krishnan2012energetic}, 
a new cathode material has been proposed and analyzed here following
up our earlier work in Ref. \cite{nemeth2018simultaneous}. The new cathode material is based
on (BN)$_{2}$OBX$_{3}$ (X is halide, such as F or Cl) 
and it may be used with Li, Na or Mg anodes forming discharge products
A$_{n}$[(BN)$_{2}$OBX$_{3}$], where A is an alkali atom (Li, Na, etc; 0$\le$n$\le$2) 
or alkaline earth (Mg, etc; 0$\le$n$\le$1).
The discharged state with one Li, Na or Mg per formula unit may be synthesized
by the thermal splitting of the Lewis adduct of the respective peroxide, such as Li$_{2}$O$_{2}$,
with BX$_{3}$, and the resulting $\cdot$OBX$_{3}^{-}$ radical anions functionalize the h-BN
monolayers in the presence of their respective cations. The radical anion functionalization method
can be extended to other 2D materials as well, to tune their properties, especially ionic and
electronic conductivities.
A$_{n}$[(BN)$_{2}$OBX$_{3}$] is predicted to
have much greater thermal stability than graphene oxide while providing similarly great energy
and power densities. It is also predicted to have good electronic and excellent ionic conductivity
and can potentially be used simultaneously as a cathode active species and a solid
electrolyte. Additional applications may include coating for metallic Li, Na
and Mg anodes, separators and polymer composites. 
The manufacturing of the proposed materials appears to be economic.

\section{Acknowledgements}
The author thanks Prof. Leon Shaw (IIT) for calling his attention to the similarity of
electronic conductivity values of (BN)$_{2}$OSO$_{2}$F (first synthesized 1978 
\cite{bartlett1978novel,shen1999intercalation})
and that of practical graphene and conductive carbon additives used in batteries.
The use of computational resources at NERSC (U.S. DOE DE-AC02-05CH11231), 
a U.S. Department of Energy Office of Science User Facility, is gratefully
acknowledged.


\begin{thebibliography}{100}

\bibitem{whittingham1976electrical}
M~Stanley Whittingham.
\newblock Electrical energy storage and intercalation chemistry.
\newblock {\em Science}, 192(4244):1126--1127, 1976.

\bibitem{godshall1980thermodynamic}
NA~Godshall, ID~Raistrick, and RA~Huggins.
\newblock Thermodynamic investigations of ternary lithium-transition
  metal-oxygen cathode materials.
\newblock {\em Materials Research Bulletin}, 15(5):561--570, 1980.

\bibitem{KMizusima80}
K.~Mizushima, P.C. Jones, P.J. Wiseman, and J.B. Goodenough.
\newblock Li$_{x}$coo$_{2}$ ($0<x<1$): A new cathode material for batteries of
  high energy density.
\newblock {\em Materials Res. Bull.}, 15:783, 1980.

\bibitem{MMThackeray01A}
Michael~M. Thackeray, Christopher~S. Johnson, Khalil Amine, and Jaekook Kim.
\newblock {Lithium metal oxide electrodes for lithium cells and batteries},
  2001.
\newblock {{US Patent} 6677082B2}.

\bibitem{MMThackeray01B}
Michael~M. Thackeray, Christopher~S. Johnson, Khalil Amine, and Jaekook Kim.
\newblock {Lithium metal oxide electrodes for lithium cells and batteries},
  2001.
\newblock {{US Patent} 6680143B2}.

\bibitem{JRDahn01}
Zhonghua Lu and Jeffrey~R. Dahn.
\newblock {Cathode compositions for lithium-ion batteries}, 2001.
\newblock {{US Patent} 6964828B2}.

\bibitem{chen2004aluminum}
CH~Chen, J~Liu, ME~Stoll, G~Henriksen, DR~Vissers, and K~Amine.
\newblock Aluminum-doped lithium nickel cobalt oxide electrodes for high-power
  lithium-ion batteries.
\newblock {\em Journal of power Sources}, 128(2):278--285, 2004.

\bibitem{AYoshino86}
Akira Yoshino, Kenichi Sanechika, and Takayuki Nakajima.
\newblock {Secondary battery}, 1986.
\newblock {{US Patent} 4668595A}.

\bibitem{RYazami83}
R.~Yazami and Ph. Touzain.
\newblock A reversible graphite-lithium negative electrode for electrochemical
  generators.
\newblock {\em J. Power Sources}, 9:365, 1983.

\bibitem{chen2014recent}
Lin Chen and Leon~L Shaw.
\newblock Recent advances in lithium--sulfur batteries.
\newblock {\em Journal of Power Sources}, 267:770--783, 2014.

\bibitem{pang2019lightweight}
Quan Pang, Chun~Yuen Kwok, Dipan Kundu, Xiao Liang, and Linda~F Nazar.
\newblock Lightweight metallic mgb2 mediates polysulfide redox and promises
  high-energy-density lithium-sulfur batteries.
\newblock {\em Joule}, 3(1):136--148, 2019.

\bibitem{ashuri2016silicon}
Maziar Ashuri, Qianran He, and Leon~L Shaw.
\newblock Silicon as a potential anode material for li-ion batteries: where
  size, geometry and structure matter.
\newblock {\em Nanoscale}, 8(1):74--103, 2016.

\bibitem{nemeth2015functionalized}
K.~Nemeth.
\newblock Functionalized boron nitride materials as electroactive species in
  electrochemical energy storage devices, 2015.
\newblock WO/2015/006161.

\bibitem{nemeth2020-US10693137}
Karoly Nemeth.
\newblock Functionalized boron nitride materials as electroactive species in
  electrochemical energy storage devices, 2020.
\newblock {US} Patent 10,693,137.

\bibitem{nemeth2014materials}
K{\'a}roly N{\'e}meth.
\newblock Materials design by quantum-chemical and other
  theoretical/computational means: Applications to energy storage and
  photoemissive materials.
\newblock {\em International Journal of Quantum Chemistry}, 114(16):1031--1035,
  2014.

\bibitem{nemeth2018simultaneous}
K{\'a}roly N{\'e}meth.
\newblock Simultaneous oxygen and boron trifluoride functionalization of
  hexagonal boron nitride: a designer cathode material for energy storage.
\newblock {\em Theoretical Chemistry Accounts}, 137(11):157, 2018.

\bibitem{bartlett1978novel}
Neil Bartlett, RN~Biagioni, BW~McQuillan, AS~Robertson, and AC~Thompson.
\newblock Novel salts of graphite and a boron nitride salt.
\newblock {\em Journal of the Chemical Society, Chemical Communications},
  5:200--201, 1978.

\bibitem{shen1999intercalation}
Ciping Shen, Steven~G Mayorga, Richard Biagioni, Charles Piskoti, Masahiro
  Ishigami, Alexander Zettl, and Neil Bartlett.
\newblock Intercalation of hexagonal boron nitride by strong oxidizers and
  evidence for the metallic nature of the products.
\newblock {\em Journal of Solid State Chemistry}, 147(1):74--81, 1999.

\bibitem{Pellion-MgIntercalation-2013}
Robert~Ellis Doe, Craig~Michael Downie, Christopher Fischer, George~Hamilton
  Lane, Dane Morgan, Josh Nevin, Gerbrand Ceder, Kristin~Aslaug Persson, and
  David Eaglesham.
\newblock {Layered materials with improved magnesium intercalation for
  rechargeable magnesium ion cells}, 2013.
\newblock {US Patent} 9401528B2.

\bibitem{Yao-MgIntercalation-2016}
Yan Yao and Hyun~Deog Yoo.
\newblock {Method of activating two-dimensional materials for
  multivalent/polyatomic-ion intercalation battery electrodes}, 2018.
\newblock {US} Patent appl. US20180183038A1.

\bibitem{yoo2017fast}
Hyun~Deog Yoo, Yanliang Liang, Hui Dong, Junhao Lin, Hua Wang, Yisheng Liu,
  Lu~Ma, Tianpin Wu, Yifei Li, Qiang Ru, et~al.
\newblock Fast kinetics of magnesium monochloride cations in
  interlayer-expanded titanium disulfide for magnesium rechargeable batteries.
\newblock {\em Nature communications}, 8(1):339, 2017.

\bibitem{Zimmerman2015}
Michael~A. Zimmerman, Randy Leising, Alexei~B. Gavrilov, Keith Smith, and Andy
  Teoli.
\newblock High capacity polymer cathode and high energy density rechargeable
  cell comprising the cathode, 2015.
\newblock {US Patent appl.} 0280218A1.

\bibitem{Zimmerman2017}
Michael~A. Zimmerman.
\newblock Solid electrolyte high energy battery, 2017.
\newblock {US Patent} 9819053B1.

\bibitem{ergen2018}
O.~Ergen and A.K. Zettl.
\newblock High temperature li-ion battery cells utilizing boron nitride
  aerogels and boron nitride nanotubes, 2018.
\newblock {US Patent} appl. 0159180.

\bibitem{rodrigues2016hexagonal}
Marco-Tulio~F Rodrigues, Kaushik Kalaga, Hemtej Gullapalli, Ganguli Babu, Arava
  Leela~Mohana Reddy, and Pulickel~M Ajayan.
\newblock Hexagonal boron nitride-based electrolyte composite for li-ion
  battery operation from room temperature to 150° c.
\newblock {\em Advanced Energy Materials}, 6(12):1600218, 2016.

\bibitem{cheng2019stabilizing}
Qian Cheng, Aijun Li, Na~Li, Shuang Li, Amirali Zangiabadi, Wenlong Huang,
  Alex~Ceng Li, Tianwei Jin, Qingquan Song, Weiheng Xu, et~al.
\newblock Stabilizing solid electrolyte-anode interface in li-metal batteries
  by boron nitride-based nanocomposite coating.
\newblock {\em Joule}, 3(6):1510--1522, 2019.

\bibitem{shen2019chemically}
Bao Shen, Tian-Wen Zhang, Yi-Chen Yin, Zheng-Xin Zhu, Lei-Lei Lu, Cheng Ma, Fei
  Zhou, and Hong-Bin Yao.
\newblock Chemically exfoliated boron nitride nanosheets form robust
  interfacial layers for stable solid-state li metal batteries.
\newblock {\em Chemical Communications}, 55(53):7703--7706, 2019.

\bibitem{rahman2019high}
Md~Mokhlesur Rahman, Srikanth Mateti, Qiran Cai, Irin Sultana, Ye~Fan, Xinwei
  Wang, Chunping Hou, and Ying Chen.
\newblock High temperature and high rate lithium-ion batteries with boron
  nitride nanotubes coated polypropylene separators.
\newblock {\em Energy Storage Materials}, 19:352--359, 2019.

\bibitem{staudenmaier1898verfahren}
L~Staudenmaier.
\newblock Verfahren zur darstellung der graphits{\"a}ure.
\newblock {\em Berichte der deutschen chemischen Gesellschaft},
  31(2):1481--1487, 1898.

\bibitem{brodie1859xiii}
Benjamin~Collins Brodie.
\newblock Xiii. on the atomic weight of graphite.
\newblock {\em Philosophical Transactions of the Royal Society of London},
  149:249--259, 1859.

\bibitem{hummers1958preparation}
William~S Hummers~Jr and Richard~E Offeman.
\newblock Preparation of graphitic oxide.
\newblock {\em Journal of the american chemical society}, 80(6):1339--1339,
  1958.

\bibitem{szabo2006evolution}
Tam{\'a}s Szab{\'o}, Ott{\'o} Berkesi, P{\'e}ter Forg{\'o}, Katalin Josepovits,
  Yiannis Sanakis, Dimitris Petridis, and Imre D{\'e}k{\'a}ny.
\newblock Evolution of surface functional groups in a series of progressively
  oxidized graphite oxides.
\newblock {\em Chemistry of materials}, 18(11):2740--2749, 2006.

\bibitem{dimiev2016graphene}
A.M. Dimiev and S.~Eigler.
\newblock {\em Graphene Oxide: Fundamentals and Applications}.
\newblock Wiley, 2016.

\bibitem{dreyer2010chemistry}
Daniel~R Dreyer, Sungjin Park, Christopher~W Bielawski, and Rodney~S Ruoff.
\newblock The chemistry of graphene oxide.
\newblock {\em Chemical society reviews}, 39(1):228--240, 2010.

\bibitem{kovtyukhova2013reversible}
Nina~I Kovtyukhova, Yuanxi Wang, Ruitao Lv, Mauricio Terrones, Vincent~H
  Crespi, and Thomas~E Mallouk.
\newblock Reversible intercalation of hexagonal boron nitride with br{\o}nsted
  acids.
\newblock {\em Journal of the American Chemical Society}, 135(22):8372--8381,
  2013.

\bibitem{jang2011graphene}
Bor~Z Jang, Chenguang Liu, David Neff, Zhenning Yu, Ming~C Wang, Wei Xiong, and
  Aruna Zhamu.
\newblock Graphene surface-enabled lithium ion-exchanging cells:
  next-generation high-power energy storage devices.
\newblock {\em Nano Letters}, 11(9):3785--3791, 2011.

\bibitem{liu2014lithium}
C.~Liu, A.~Zhamu, D.~Neff, and B.Z. Jang.
\newblock Lithium super-battery with a functionalized nano graphene cathode,
  2014.
\newblock {US Patent} 8795899B2.

\bibitem{kim2014novel}
Haegyeom Kim, Young-Uk Park, Kyu-Young Park, Hee-Dae Lim, Jihyun Hong, and
  Kisuk Kang.
\newblock Novel transition-metal-free cathode for high energy and power sodium
  rechargeable batteries.
\newblock {\em Nano Energy}, 4:97--104, 2014.

\bibitem{kim2014all}
Haegyeom Kim, Kyu-Young Park, Jihyun Hong, and Kisuk Kang.
\newblock All-graphene-battery: bridging the gap between supercapacitors and
  lithium ion batteries.
\newblock {\em Scientific reports}, 4:5278, 2014.

\bibitem{petrenko1902}
G.~I. Petrenko.
\newblock Derivatives of perboric acid.
\newblock {\em J. Russ. Phys. Chem. Soc.}, 34:37--42, 1902.
\newblock https://books.google.com/books?id=7fU4AAAAMAAJ, page 317.

\bibitem{samanta2013highly}
Khokan Samanta, Surajit Some, Youngmin Kim, Yeoheung Yoon, Misook Min, Sae~Mi
  Lee, Younghun Park, and Hyoyoung Lee.
\newblock Highly hydrophilic and insulating fluorinated reduced graphene oxide.
\newblock {\em Chemical Communications}, 49(79):8991--8993, 2013.

\bibitem{koga2018multistep}
Nobuyoshi Koga, Nao Kameno, Yoji Tsuboi, Takayuki Fujiwara, Masayoshi Nakano,
  Kazuyuki Nishikawa, and Akiko~Iwasaki Murata.
\newblock Multistep thermal decomposition of granular sodium perborate
  tetrahydrate: A kinetic approach to complex reactions in solid--gas systems.
\newblock {\em Physical Chemistry Chemical Physics}, 20(18):12557--12573, 2018.

\bibitem{BF3Al2O3Patent1947}
Charles~Joseph Plank.
\newblock Catalytic conversion of hydrocarbons, 1947.
\newblock {US Patent} US2428741A.

\bibitem{rhee1968chemisorption}
Kee~H Rhee and Michael~R Basila.
\newblock Chemisorption of bf3 on catalytic oxide surfaces: Infrared
  spectroscopic studies.
\newblock {\em Journal of Catalysis}, 10(3):243--251, 1968.

\bibitem{morrow1976infrared}
BA~Morrow, IA~Cody, and Lydia~SM Lee.
\newblock Infrared studies of reactions on oxide surfaces. 7. mechanism of the
  adsorption of water and ammonia on dehydroxylated silica.
\newblock {\em The Journal of Physical Chemistry}, 80(25):2761--2767, 1976.

\bibitem{sadeghi2008bf}
BFMB Sadeghi, BF~Mirjalili, and MM~Hashemi.
\newblock Bf 3. sio 2: An efficient heterogeneous alternative for regio-chemo
  and stereoselective claisen-schmidt condensation.
\newblock {\em Journal of the Iranian Chemical Society}, 5(4):694--698, 2008.

\bibitem{nie2016some}
Mengyun Nie, L~Madec, J~Xia, DS~Hall, and JR~Dahn.
\newblock Some lewis acid-base adducts involving boron trifluoride as
  electrolyte additives for lithium ion cells.
\newblock {\em Journal of Power Sources}, 328:433--442, 2016.

\bibitem{xiao2016lewisadducts}
Ang Xiao, William~M. Lamanna, Jeffrey~R. Dahn, Mengyun Nie, Kiah~A. Smith, and
  Vincent~J. Chevrier.
\newblock Electrochemical cells that include lewis acid: lewis base complex
  electrolyte additives, 2016.
\newblock Patent appl. WO/2016/126534A1.

\bibitem{matsui2014}
Masaki Matsui.
\newblock Asymmetric type {BF$_{3}$} complex, 2014.
\newblock {US Patent} 8765294B2.

\bibitem{wang2000}
Yu~Wang, Meijie Zhang, Ulrich~Von Sacken, and Brian~Michael Way.
\newblock Additives for improving cycle life of non-aqueous rechargeable
  lithium batteries, 2000.
\newblock {US} Patent US/6045948A.

\bibitem{carino2015bf}
Emily~V Carino, Charles~E Diesendruck, Jeffrey~S Moore, Larry~A Curtiss,
  Rajeev~S Assary, and Fikile~R Brushett.
\newblock Bf 3-promoted electrochemical properties of quinoxaline in propylene
  carbonate.
\newblock {\em RSC Advances}, 5(24):18822--18831, 2015.

\bibitem{ye2017metal}
Minghui Ye, Jian Gao, Yukun Xiao, Tong Xu, Yang Zhao, and Liangti Qu.
\newblock Metal/graphene oxide batteries.
\newblock {\em Carbon}, 125:299--307, 2017.

\bibitem{venkateswarlu2019effective}
Gundu Venkateswarlu, Devarapaga Madhu, and Jetti~Vatsala Rani.
\newblock An effective performance of f-doped hexagonal boron nitride
  nanosheets as cathode material in magnesium battery.
\newblock {\em Materials Chemistry and Physics}, 226:356--361, 2019.

\bibitem{CFxBattery1970}
Nobuatsu Watanabe and Masataro Fukuda.
\newblock Primary cell for electric batteries, 1970.
\newblock {US Patent} 3536532A.

\bibitem{entani2020non}
Shiro Entani, Konstantin~V Larionov, Zakhar~I Popov, Masaru Takizawa, Masaki
  Mizuguchi, Hideo Watanabe, Songtian Li, Hiroshi Naramoto, Pavel~B Sorokin,
  and Seiji Sakai.
\newblock Non-chemical fluorination of hexagonal boron nitride by high-energy
  ion irradiation.
\newblock {\em Nanotechnology}, 31(12):125705, 2020.

\bibitem{kim2010self}
Franklin Kim, Jiayan Luo, Rodolfo Cruz-Silva, Laura~J Cote, Kwonnam Sohn, and
  Jiaxing Huang.
\newblock Self-propagating domino-like reactions in oxidized graphite.
\newblock {\em Advanced Functional Materials}, 20(17):2867--2873, 2010.

\bibitem{krishnan2012energetic}
Deepti Krishnan, Franklin Kim, Jiayan Luo, Rodolfo Cruz-Silva, Laura~J Cote,
  Hee~Dong Jang, and Jiaxing Huang.
\newblock Energetic graphene oxide: challenges and opportunities.
\newblock {\em Nano today}, 7(2):137--152, 2012.

\bibitem{lowe2017challenges}
Sean~E Lowe and Yu~Lin Zhong.
\newblock Challenges of industrial-scale graphene oxide production.
\newblock {\em Graphene Oxide: Fundamentals and Applications; Dimiev AM, Eigler
  S., Eds}, pages 410--431, 2017.

\bibitem{lin2019synthesis}
Li~Lin, Hailin Peng, and Zhongfan Liu.
\newblock Synthesis challenges for graphene industry.
\newblock {\em Nature materials}, 18(6):520--524, 2019.

\bibitem{ikuno2007amine}
T~Ikuno, T~Sainsbury, D~Okawa, JMJ Fr{\'e}chet, and A~Zettl.
\newblock Amine-functionalized boron nitride nanotubes.
\newblock {\em Solid State Communications}, 142(11):643--646, 2007.

\bibitem{nazarov2012functionalization}
Albert~S Nazarov, Viktor~N Demin, Ekaterina~D Grayfer, Alexander~I Bulavchenko,
  Aida~T Arymbaeva, Hyeon-Jin Shin, Jae-Young Choi, and Vladimir~E Fedorov.
\newblock Functionalization and dispersion of hexagonal boron nitride (h-bn)
  nanosheets treated with inorganic reagents.
\newblock {\em Chemistry--An Asian Journal}, 7(3):554--560, 2012.

\bibitem{RU2478077C2}
V.E. Fedorov, A.S. Nazarov, and V.N. Demin.
\newblock Method of producing soluble hexagonal boron nitride, 2013.
\newblock Russian Patent RU 2478077 C2.

\bibitem{pakdel2014plasma}
Amir Pakdel, Yoshio Bando, and Dmitri Golberg.
\newblock Plasma-assisted interface engineering of boron nitride nanostructure
  films.
\newblock {\em ACS nano}, 8(10):10631--10639, 2014.

\bibitem{cui2014large}
Zhenhua Cui, Andrew~J Oyer, A~Jaeton Glover, Hannes~C Schniepp, and Douglas~H
  Adamson.
\newblock Large scale thermal exfoliation and functionalization of boron
  nitride.
\newblock {\em Small}, 10(12):2352--2355, 2014.

\bibitem{li2014strong}
Lu~Hua Li, Jiri Cervenka, Kenji Watanabe, Takashi Taniguchi, and Ying Chen.
\newblock Strong oxidation resistance of atomically thin boron nitride
  nanosheets.
\newblock {\em ACS nano}, 8(2):1457--1462, 2014.

\bibitem{anota2011lda}
E~Chigo Anota, H~Hern{\'a}ndez Cocoletzi, and E~Rubio Rosas.
\newblock Lda approximation based analysis of the adsorption of o3 by boron
  nitride sheet.
\newblock {\em The European Physical Journal D}, 63(2):271--273, 2011.

\bibitem{nemeth2017OzoneBF3}
K.~Nemeth, A.~M. Danby, and B.~Subramaniam.
\newblock Synthesis of oxygen and boron trihalogenide functionalized
  two-dimensional layered materials in pressurized medium, 2017.
\newblock Patent appl. WO/2017/196738 and US/2019/0134585.

\bibitem{sainsbury2012oxygen}
Toby Sainsbury, Amro Satti, Peter May, Zhiming Wang, Ignatius McGovern, Yurii~K
  Gun’ko, and Jonathan Coleman.
\newblock Oxygen radical functionalization of boron nitride nanosheets.
\newblock {\em Journal of the American Chemical Society}, 134(45):18758--18771,
  2012.

\bibitem{mofakhami2010material}
A.~Mofakhami and J.F. Fauvarque.
\newblock Method of activating boron nitride, 2010.
\newblock {US Patent} appl. 2010/0280138A1.

\bibitem{mofakhami2011material}
A.~Mofakhami and J.F. Fauvarque.
\newblock Material for an electrochemical device, 2011.
\newblock {US Patent} appl. 2011/0091789A1.

\bibitem{mofakhami2015material}
A.~Mofakhami and J.F. Fauvarque.
\newblock Material for an electrochemical device, 2015.
\newblock {US} Patent 9,105,907B2.

\bibitem{rajendran2012surface}
G.P. Rajendran.
\newblock Surface modified hexagonal boron nitride particles, 2012.
\newblock {US Patent} 8258346.

\bibitem{sheng2019polymer}
Wenbo Sheng, Ihsan Amin, Christof Neumann, Renhao Dong, Tao Zhang, Erik
  Wegener, Wei-Liang Chen, Paul F{\"o}rster, Hai~Quang Tran, Markus
  L{\"o}ffler, et~al.
\newblock Polymer brushes: Polymer brushes on hexagonal boron nitride (small
  19/2019).
\newblock {\em Small}, 15(19):1970099, 2019.

\bibitem{sainsbury2014dibromocarbene}
Toby Sainsbury, Arlene O’Neill, Melissa~K Passarelli, Maud Seraffon, Dipak
  Gohil, Sam Gnaniah, Steve~J Spencer, Alasdair Rae, and Jonathan~N Coleman.
\newblock Dibromocarbene functionalization of boron nitride nanosheets: Toward
  band gap manipulation and nanocomposite applications.
\newblock {\em Chemistry of Materials}, 26(24):7039--7050, 2014.

\bibitem{lin2009soluble}
Yi~Lin, Tiffany~V Williams, and John~W Connell.
\newblock Soluble, exfoliated hexagonal boron nitride nanosheets.
\newblock {\em The Journal of Physical Chemistry Letters}, 1(1):277--283, 2009.

\bibitem{lee2015scalable}
Dongju Lee, Bin Lee, Kwang~Hyun Park, Ho~Jin Ryu, Seokwoo Jeon, and Soon~Hyung
  Hong.
\newblock Scalable exfoliation process for highly soluble boron nitride
  nanoplatelets by hydroxide-assisted ball milling.
\newblock {\em Nano letters}, 15(2):1238--1244, 2015.

\bibitem{zhang2016experimental}
Fan Zhang, K{\'a}roly N{\'e}meth, Javier Bare{\~n}o, Fulya Dogan, Ira~D Bloom,
  and Leon~L Shaw.
\newblock Experimental and theoretical investigations of functionalized boron
  nitride as electrode materials for li-ion batteries.
\newblock {\em RSC Advances}, 6(33):27901--27914, 2016.

\bibitem{wentorf1961synthesis}
RH~Wentorf~Jr.
\newblock Synthesis of the cubic form of boron nitride.
\newblock {\em The Journal of Chemical Physics}, 34(3):809--812, 1961.

\bibitem{bakandritsos2017cyanographene}
Aristides Bakandritsos, Martin Pykal, Piotr B{\l}oński, Petr Jakubec,
  Demetrios~D Chronopoulos, Kateřina Poláková, Vasilios Georgakilas,
  Klára Čépe, Ondřej Tomanec, Václav Ranc, Athanasios~B. Bourlinos,
  Radek Zbořil, and Michal Otyepka.
\newblock Cyanographene and graphene acid: emerging derivatives enabling
  high-yield and selective functionalization of graphene.
\newblock {\em ACS nano}, 11(3):2982--2991, 2017.

\bibitem{cheong2019cyanographene}
Yi~Heng Cheong, Muhammad Zafir~Mohamad Nasir, Aristides Bakandritsos, Martin
  Pykal, Petr Jakubec, Radek Zbo{\v{r}}il, Michal Otyepka, and Martin Pumera.
\newblock Cyanographene and graphene acid: the functional group of graphene
  derivative determines the application in electrochemical sensing and
  capacitors.
\newblock {\em ChemElectroChem}, 6(1):229--234, 2019.

\bibitem{talande2019densely}
Smita~V Talande, Aristides Bakandritsos, Petr Jakubec, Ond{\v{r}}ej Malina,
  Radek Zbo{\v{r}}il, and Ji{\v{r}}i Tu{\v{c}}ek.
\newblock Densely functionalized cyanographene bypasses aqueous electrolytes
  and synthetic limitations toward seamless graphene/$\beta$-feooh hybrids for
  supercapacitors.
\newblock {\em Advanced Functional Materials}, 29(51):1906998, 2019.

\bibitem{jeong2011nitrogen}
Hyung~Mo Jeong, Jung~Woo Lee, Weon~Ho Shin, Yoon~Jeong Choi, Hyun~Joon Shin,
  Jeung~Ku Kang, and Jang~Wook Choi.
\newblock Nitrogen-doped graphene for high-performance ultracapacitors and the
  importance of nitrogen-doped sites at basal planes.
\newblock {\em Nano letters}, 11(6):2472--2477, 2011.

\bibitem{emani2019li3bn2}
Satyanarayana Emani, Caihong Liu, Maziar Ashuri, Karan Sahni, Jinpeng Wu, Wanli
  Yang, K{\'a}roly N{\'e}meth, and Leon~L Shaw.
\newblock Li3bn2 as a transition metal free, high capacity cathode for li-ion
  batteries.
\newblock {\em ChemElectroChem}, 6(2):320--325, 2019.

\bibitem{sainsbury2016covalent}
Toby Sainsbury, Melissa Passarelli, Mira Naftaly, Sam Gnaniah, Steve~J Spencer,
  and Andrew~J Pollard.
\newblock Covalent carbene functionalization of graphene: Toward chemical
  band-gap manipulation.
\newblock {\em ACS applied materials \& interfaces}, 8(7):4870--4877, 2016.

\bibitem{rani2010electrical}
Adila Rani, Seung-Woong Nam, Kyoung-Ah Oh, and Min Park.
\newblock Electrical conductivity of chemically reduced graphene powders under
  compression.
\newblock {\em Carbon Letters (Carbon Lett.)}, 11(2):90--95, 2010.

\bibitem{pantea2003electrical}
Dana Pantea, Hans Darmstadt, Serge Kaliaguine, and Christian Roy.
\newblock Electrical conductivity of conductive carbon blacks: influence of
  surface chemistry and topology.
\newblock {\em Applied Surface Science}, 217(1-4):181--193, 2003.

\bibitem{pervez2019interface}
Syed~Atif Pervez, Musa~Ali Cambaz, Venkataraman Thangadurai, and Maximilian
  Fichtner.
\newblock Interface in solid-state lithium battery: Challenges, progress, and
  outlook.
\newblock {\em ACS applied materials \& interfaces}, 11(25):22029--22050, 2019.

\bibitem{mattson2011evidence}
Eric~C Mattson, Haihui Pu, Shumao Cui, Marvin~A Schofield, Sonny Rhim, Ganhua
  Lu, Michael~J Nasse, Rodney~S Ruoff, Michael Weinert, Marija
  Gajdardziska-Josifovska, et~al.
\newblock Evidence of nanocrystalline semiconducting graphene monoxide during
  thermal reduction of graphene oxide in vacuum.
\newblock {\em ACS nano}, 5(12):9710--9717, 2011.

\bibitem{zhou2013origin}
Si~Zhou and Angelo Bongiorno.
\newblock Origin of the chemical and kinetic stability of graphene oxide.
\newblock {\em Scientific reports}, 3:2484, 2013.

\bibitem{devilliers1983mass}
Didier Devilliers, Michel Vogler, Frederic Lantelme, and Marius Chemla.
\newblock Mass spectrometric analysis of thermal decomposition products of
  graphite fluorides and electrogenerated carbon—fluorine compounds.
\newblock {\em Analytica chimica acta}, 153:69--82, 1983.

\bibitem{GPacchioni05}
Gianfranco Pacchioni, Livia Giordano, and Matteo Baistrocchi.
\newblock Charging of metal atoms on ultrathin mgo/mo(100) films.
\newblock {\em Phys. Rev. Lett.}, 94:226104, 2005.

\bibitem{QE-2009}
Paolo Giannozzi, Stefano Baroni, Nicola Bonini, Matteo Calandra, Roberto Car,
  Carlo Cavazzoni, Davide Ceresoli, Guido~L Chiarotti, Matteo Cococcioni,
  Ismaila Dabo, Andrea {Dal Corso}, Stefano de~Gironcoli, Stefano Fabris, Guido
  Fratesi, Ralph Gebauer, Uwe Gerstmann, Christos Gougoussis, Anton Kokalj,
  Michele Lazzeri, Layla Martin-Samos, Nicola Marzari, Francesco Mauri,
  Riccardo Mazzarello, Stefano Paolini, Alfredo Pasquarello, Lorenzo Paulatto,
  Carlo Sbraccia, Sandro Scandolo, Gabriele Sclauzero, Ari~P Seitsonen,
  Alexander Smogunov, Paolo Umari, and Renata~M Wentzcovitch.
\newblock Quantum espresso: a modular and open-source software project for
  quantum simulations of materials.
\newblock {\em Journal of Physics: Condensed Matter}, 21(39):395502 (19pp),
  2009.

\bibitem{QE-2017}
P~Giannozzi, O~Andreussi, T~Brumme, O~Bunau, M~Buongiorno Nardelli, M~Calandra,
  R~Car, C~Cavazzoni, D~Ceresoli, M~Cococcioni, N~Colonna, I~Carnimeo, A~Dal
  Corso, S~de~Gironcoli, P~Delugas, R~A~DiStasio Jr, A~Ferretti, A~Floris,
  G~Fratesi, G~Fugallo, R~Gebauer, U~Gerstmann, F~Giustino, T~Gorni, J~Jia,
  M~Kawamura, H-Y Ko, A~Kokalj, E~Küçükbenli, M~Lazzeri, M~Marsili,
  N~Marzari, F~Mauri, N~L Nguyen, H-V Nguyen, A~Otero de-la Roza, L~Paulatto,
  S~Poncé, D~Rocca, R~Sabatini, B~Santra, M~Schlipf, A~P Seitsonen,
  A~Smogunov, I~Timrov, T~Thonhauser, P~Umari, N~Vast, X~Wu, and S~Baroni.
\newblock Advanced capabilities for materials modelling with quantum espresso.
\newblock {\em Journal of Physics: Condensed Matter}, 29(46):465901, 2017.

\bibitem{nemeth2014ultrahigh}
K{\'a}roly N{\'e}meth.
\newblock Ultrahigh energy density li-ion batteries based on cathodes of 1d
  metals with--li--n--b--n--repeating units in $\alpha$-lixbn2 (1⩽ x⩽ 3).
\newblock {\em The Journal of chemical physics}, 141(5):054711, 2014.

\bibitem{PBE}
John~P. Perdew, Kieron Burke, and Matthias Ernzerhof.
\newblock Generalized gradient approximation made simple.
\newblock {\em Phys. Rev. Lett.}, 77:3865, 1996.

\bibitem{PBEsol}
John~P. Perdew, Adrienn Ruzsinszky, G{\'a}bor~I. Csonka, O.~A. Vydrov, G.~E.
  Scuseria, L.~A. Constantin, X.~Zhou, and K.~Burke.
\newblock Restoring the density-gradient expansion for exchange in solids and
  surfaces.
\newblock {\em Phys. Rev. Lett.}, 100:136406, 2008.

\bibitem{zolyomi2015towards}
Viktor Zolyomi and J~K{\"u}rti.
\newblock Towards improved exact exchange functionals relying on g w
  quasiparticle methods for parametrization.
\newblock {\em Physical Review B}, 92(3):035150, 2015.

\bibitem{PBE01perdew1996rationale}
John~P Perdew, Matthias Ernzerhof, and Kieron Burke.
\newblock Rationale for mixing exact exchange with density functional
  approximations.
\newblock {\em The Journal of chemical physics}, 105(22):9982--9985, 1996.

\bibitem{PBE02adamo1999toward}
Carlo Adamo and Vincenzo Barone.
\newblock Toward reliable density functional methods without adjustable
  parameters: The pbe0 model.
\newblock {\em The Journal of chemical physics}, 110(13):6158--6170, 1999.

\bibitem{heyd2003hybrid}
Jochen Heyd, Gustavo~E Scuseria, and Matthias Ernzerhof.
\newblock Hybrid functionals based on a screened coulomb potential.
\newblock {\em The Journal of chemical physics}, 118(18):8207--8215, 2003.

\bibitem{watanabe2004direct}
Kenji Watanabe, Takashi Taniguchi, and Hisao Kanda.
\newblock Direct-bandgap properties and evidence for ultraviolet lasing of
  hexagonal boron nitride single crystal.
\newblock {\em Nature materials}, 3(6):404--409, 2004.

\bibitem{mofakhamipatents}
A.~Mofakhami and J.F. Fauvarque.
\newblock Material for an electrochemical device, 2010.
\newblock {US Patent appls.} 2010/0280138, 2011/0091789A1, 9105907B2.

\bibitem{lu2006thermal}
Zhenrong Lu, Li~Yang, and Yaju Guo.
\newblock Thermal behavior and decomposition kinetics of six electrolyte salts
  by thermal analysis.
\newblock {\em Journal of power sources}, 156(2):555--559, 2006.

\bibitem{polishchuk2011oxyfluoride}
SA~Polishchuk, LN~Ignat’eva, Yu~V Marchenko, and VM~Bouznik.
\newblock Oxyfluoride glasses (a review).
\newblock {\em Glass Physics and Chemistry}, 37(1):1--20, 2011.

\bibitem{boussard1997structure}
Catherine Boussard-Pl{\'e}del, Marie Le~Floch, Gilles Fonteneau, Jacques Lucas,
  Sourisak Sinbandhit, J~Shao, CA~Angell, Jo{\"e}l Emery, and JY~Buzare.
\newblock The structure of a boron oxyfluoride glass, an inorganic cross-linked
  chain polymer.
\newblock {\em Journal of non-crystalline solids}, 209(3):247--256, 1997.

\bibitem{ozonolysisBala2013}
Bala Subramaniam, Daryle Busch, Andrew~M. Danby, and Thomas~P Binder.
\newblock Ozonolysis reactions in liquid {CO$_{2}$} and {CO$_{2}$}-expanded
  solvents, 2013.
\newblock {US} Patent US8801939B2.

\bibitem{lundin2015liquid}
Michael~D Lundin, Andrew~M Danby, Geoffrey~R Akien, Thomas~P Binder, Daryle~H
  Busch, and Bala Subramaniam.
\newblock Liquid co2 as a safe and benign solvent for the ozonolysis of fatty
  acid methyl esters.
\newblock {\em ACS Sustainable Chemistry \& Engineering}, 3(12):3307--3314,
  2015.

\bibitem{mcclure1962hydrogen}
James~D McClure and Paul~H Williams.
\newblock Hydrogen peroxide—boron trifluoride etherate, a new oxidizing
  agent.
\newblock {\em The Journal of Organic Chemistry}, 27(1):24--26, 1962.

\bibitem{nemeth2019radicalAnion}
Karoly Nemeth.
\newblock Radical anion functionalization of two-dimensional materials, 2020.
\newblock {Patent appl.} PCT/US/20/20383, WO/2020/180680.

\bibitem{hartman1978adducts}
J~Stephen Hartman and Jack~M Miller.
\newblock Adducts of the mixed trihalides of boron.
\newblock In {\em Advances in Inorganic Chemistry and Radiochemistry},
  volume~21, pages 147--177. Elsevier, 1978.

\bibitem{Gilpatrick1974}
Louis~O. Gilpatrick.
\newblock Synthesis of sodium hydroxytri-fluoroborate, 1974.
\newblock {US Patent} 3809762.

\bibitem{clark1970crystal}
MJR Clark and H~Lynton.
\newblock The crystal and molecular structure of nabf3oh.
\newblock {\em Canadian Journal of Chemistry}, 48(3):405--409, 1970.

\bibitem{Zhang2015KineticSO}
Weijiang Zhang, Xiangmei Wang, and Jiao Xu.
\newblock Kinetic study on preparation of high-purity enriched boric-10 acid
  used in nuclear power plants.
\newblock {\em Asian Journal of Chemistry}, 27(9):3234--3238, 2015.

\bibitem{wamser1951equilibria}
Christian~A Wamser.
\newblock Equilibria in the system boron trifluoride—water at 25.
\newblock {\em Journal of the American Chemical Society}, 73(1):409--416, 1951.

\bibitem{paterson1961interaction}
WG~Paterson and M~Onyszchuk.
\newblock The interaction of boron trifluoride with hydrazine.
\newblock {\em Canadian Journal of Chemistry}, 39(5):986--994, 1961.

\bibitem{KDonggun2020adfm}
Donggun Kim, Xin Liu, Baozhi Yu, Srikanth Mateti, Luke~A. O'Dell, Qiangzhou
  Rong, and Ying~(Ian) Chen.
\newblock Amine-functionalized boron nitride nanosheets: A new functional
  additive for robust, flexible ion gel electrolyte with high lithium-ion
  transference number.
\newblock {\em Advanced Functional Materials}, 30(15):1910813, 2020.

\bibitem{zhang2020targeting}
Zhizhen Zhang, Hui Li, Kavish Kaup, Laidong Zhou, Pierre-Nicholas Roy, and
  Linda~F Nazar.
\newblock Targeting superionic conductivity by turning on anion rotation at
  room temperature in fast ion conductors.
\newblock {\em Matter}, 2(6):1667--1684, 2020.

\bibitem{yoon2002oxygen}
Won-Sub Yoon, Kwang-Bum Kim, Min-Gyu Kim, Min-Kyu Lee, Hyun-Joon Shin, Jay-Min
  Lee, Jae-Sung Lee, and Chul-Hyun Yo.
\newblock Oxygen contribution on li-ion intercalation- deintercalation in
  licoo2 investigated by o k-edge and co l-edge x-ray absorption spectroscopy.
\newblock {\em The Journal of Physical Chemistry B}, 106(10):2526--2532, 2002.

\bibitem{mizokawa2013role}
T~Mizokawa, Y~Wakisaka, T~Sudayama, C~Iwai, K~Miyoshi, J~Takeuchi, H~Wadati,
  DG~Hawthorn, TZ~Regier, and GA~Sawatzky.
\newblock Role of oxygen holes in li x coo 2 revealed by soft x-ray
  spectroscopy.
\newblock {\em Physical Review Letters}, 111(5):056404, 2013.

\bibitem{seo2016structural}
Dong-Hwa Seo, Jinhyuk Lee, Alexander Urban, Rahul Malik, ShinYoung Kang, and
  Gerbrand Ceder.
\newblock The structural and chemical origin of the oxygen redox activity in
  layered and cation-disordered li-excess cathode materials.
\newblock {\em Nature chemistry}, 8(7):692, 2016.

\bibitem{luo2016charge}
Kun Luo, Matthew~R Roberts, Rong Hao, Niccol{\'o} Guerrini, David~M Pickup,
  Yi-Sheng Liu, Kristina Edstr{\"o}m, Jinghua Guo, Alan~V Chadwick, Laurent~C
  Duda, and Peter~G. Bruce.
\newblock Charge-compensation in 3d-transition-metal-oxide intercalation
  cathodes through the generation of localized electron holes on oxygen.
\newblock {\em Nature Chemistry}, 8(7):684, 2016.

\bibitem{shen2019lithium}
Xiaowei Shen, Yutao Li, Tao Qian, Jie Liu, Jinqiu Zhou, Chenglin Yan, and
  John~B Goodenough.
\newblock Lithium anode stable in air for low-cost fabrication of a
  dendrite-free lithium battery.
\newblock {\em Nature Communications}, 10(1):1--9, 2019.

\bibitem{huang2020graphitic}
Ying Huang, Bo~Chen, Jian Duan, Fei Yang, Tengrui Wang, Zhengfeng Wang, Wenjuan
  Yang, Chenchen Hu, Wei Luo, and Yunhui Huang.
\newblock Graphitic carbon nitride (g-c3n4): An interface enabler for
  solid-state lithium metal batteries.
\newblock {\em Angewandte Chemie International Edition}, 59(9):3699--3704,
  2020.

\bibitem{ju2020biomacromolecules}
Zhijin Ju, Jianwei Nai, Yao Wang, Tiefeng Liu, Jianhui Zheng, Huadong Yuan,
  Ouwei Sheng, Chengbin Jin, Wenkui Zhang, Zhong Jin, et~al.
\newblock Biomacromolecules enabled dendrite-free lithium metal battery and its
  origin revealed by cryo-electron microscopy.
\newblock {\em Nature communications}, 11(1):1--10, 2020.

\bibitem{qian2015high}
Jiangfeng Qian, Wesley~A Henderson, Wu~Xu, Priyanka Bhattacharya, Mark
  Engelhard, Oleg Borodin, and Ji-Guang Zhang.
\newblock High rate and stable cycling of lithium metal anode.
\newblock {\em Nature communications}, 6(1):1--9, 2015.

\bibitem{shkrob2014fluorosulfonyl}
Ilya~A Shkrob, Timothy~W Marin, Ye~Zhu, and Daniel~P Abraham.
\newblock Why bis (fluorosulfonyl) imide is a “magic anion” for
  electrochemistry.
\newblock {\em The Journal of Physical Chemistry C}, 118(34):19661--19671,
  2014.

\bibitem{US7534527B2}
Zhiqiang Xu, Chi kyun Park, Zhiwei Zhang, and Chai Chul.
\newblock Organic lithium salt electrolytes having enhanced safety for
  rechargeable batteries and methods of making the same, 2009.
\newblock {US Patent} 7534527B2.

\bibitem{JP2013209355A}
Kenichi Shinmyo, Tomoe Yoshida, and Masashi Kano.
\newblock Lithium carboxylate salt-boron trifluoride complex, method for
  manufacturingthe complex, electrolyte solution, method for manufacturing the
  electrolyte solution, gel electrolyte, and solid electrolyte, 2013.
\newblock {Japanese Patent} 2013209355A.

\bibitem{JP2017178964A}
Kenichi Shinmyo and Katsu Heiji.
\newblock Process for producing lithium carboxylic acid salt-boron trifluoride
  complex, 2017.
\newblock {Japanese Patent} 2017178964A.

\bibitem{zhou2020protective}
Hongyao Zhou, Sicen Yu, Haodong Liu, and Ping Liu.
\newblock Protective coatings for lithium metal anodes: Recent progress and
  future perspectives.
\newblock {\em Journal of Power Sources}, 450:227632, 2020.

\bibitem{prices}
Prices of h-BN, BF$_{3}$ and BCl$_{3}$ are from www.alibaba.com (various
  suppliers). Prices of Li, Na and Mg are from price.metal.com (Shanghai Metals
  Market).

\bibitem{trahey2020energy}
Lynn Trahey, Fikile~R Brushett, Nitash~P Balsara, Gerbrand Ceder, Lei Cheng,
  Yet-Ming Chiang, Nathan~T Hahn, Brian~J Ingram, Shelley~D Minteer, Jeffrey~S
  Moore, et~al.
\newblock Energy storage emerging: A perspective from the joint center for
  energy storage research.
\newblock {\em Proceedings of the National Academy of Sciences},
  117(23):12550--12557, 2020.

\end{thebibliography}

\end{document}